\documentclass[
reprint,
longbibliography,
aps,
twoside,
superscriptaddress,
prb, nobibnotes,
floatfix
]{revtex4-2}
\pdfoutput=1

\usepackage{physics}
\usepackage{color,newfloat}
\usepackage{graphicx}
\usepackage{svg}
\usepackage{dcolumn}\usepackage{bm}
\usepackage{amsmath,amssymb,amsthm,amsfonts}
\usepackage{bbm}
\usepackage{siunitx}
\usepackage[vcentermath]{youngtab}
\usepackage{mathtools}
\usepackage{multirow}
\usepackage{enumerate}
\usepackage{lipsum}
\usepackage{placeins}
\usepackage[labelfont=bf]{caption}
\usepackage{algorithm}
\usepackage{algpseudocode}
\usepackage{caption}
\usepackage{ragged2e}
\usepackage{dsfont}

\usepackage{hyperref} 
\hypersetup{
    colorlinks=true,
    linkcolor=black,
    urlcolor=brown,
    linktoc=all,
    citecolor=brown
           }
\newcommand\NoDo{\renewcommand\algorithmicdo{}}
\newcommand\ReDo{\renewcommand\algorithmicdo{\textbf{do}}}
\newcommand{\MYhref}[3][black]{\href{#2}{\color{#1}{#3}}}%

\begin{document}

\title{Errors in heralded circuits for linear optical entanglement generation}

\author{Reece D. Shaw}
\email{reece.shaw@bristol.ac.uk}
\affiliation{Quantum Engineering Technology Labs, H.H Wills Physics Laboratory and Department of Electrical and Electronic Engineering, University of Bristol, Bristol, United Kingdom}
\affiliation{Quantum Engineering Centre for Doctoral Training,
Centre for Nanoscience and Quantum Information, University of Bristol, Bristol, United Kingdom}
\author{Alex E. Jones}
\affiliation{Quantum Engineering Technology Labs, H.H Wills Physics Laboratory and Department of Electrical and Electronic Engineering, University of Bristol, Bristol, United Kingdom}
\author{Patrick Yard}
\affiliation{Quantum Engineering Technology Labs, H.H Wills Physics Laboratory and Department of Electrical and Electronic Engineering, University of Bristol, Bristol, United Kingdom}
\author{Anthony Laing}
\affiliation{Quantum Engineering Technology Labs, H.H Wills Physics Laboratory and Department of Electrical and Electronic Engineering, University of Bristol, Bristol, United Kingdom}
\date{\today}


\begin{abstract}
\normalsize
The heralded generation of entangled states underpins many photonic quantum technologies. As quantum error correction thresholds are determined by underlying physical noise mechanisms, a detailed and faithful characterization of resource states is required. Non-computational leakage, e.g. more than one photon occupying a dual-rail encoded qubit, is an error not captured by standard forms of state tomography, which postselect on photons remaining in the computational subspace. Here we use the continuous-variable (CV) formalism and first quantized state representation to develop a simulation framework that reconstructs photonic quantum states in the presence of partial distinguishability and resulting non-computational leakage errors. Using these tools, we analyze a variety of Bell state generation circuits and find that the five photon discrete Fourier transform (DFT) Bell state generation scheme [\MYhref{https://journals.aps.org/prl/abstract/10.1103/PhysRevLett.126.230504}{Phys Rev. Lett. \textbf{126} 23054 (2021)}] is most robust to such errors for near-ideal photons. Through characterization of a photonic entangling gate, we demonstrate how leakage errors prevent a modular characterization of concatenated gates using current tomographical procedures. Our work is a necessary step in revealing the true noise models that must be addressed in fault-tolerant photonic quantum computing architectures.

\end{abstract}
\maketitle

The workhorse of linear optical quantum computation (LOQC) is Hong-Ou-Mandel (HOM) interference which enables indirect interactions and subsequent entanglement between single photons \cite{hong_measurement_1987}. The first proposal for universal quantum computation with photons was the Knill, Laflamme and Milburn (KLM) scheme that requires a large number of optical elements, quantum memory, and phase stability of the resulting interferometer \cite{knill_scheme_2001}. More recently, prominent proposals for the implementation of LOQC, which are built upon the measurement-based paradigm of quantum computing (MBQC), combine small photonic entangled states through fusion operations. This provides a universal computational resource on which algorithms can be executed through single-qubit measurements \cite{raussendorf_measurement-based_2003}.  The overheads of KLM are now shifted to the generation and fusion of small entangled resource states (Bell, Greenberger-Horne-Zeilinger (GHZ)) \cite{gimeno-segovia_three-photon_2015} composed of indistinguishable photons. The most modern architecture for universal photonic quantum computing, known as fusion-based quantum computation (FBQC), also relies on the generation of Bell and GHZ seed states. Seed states are consumed in resource state generators to provide states used to construct fault-tolerant fusion networks \cite{fusion-based-qc, bartolucci-2021, bombin_interleaving_2021}.

In practice, no two photons in a device will be perfectly indistinguishable and no device will operate without error. Physical errors in photonics may include photon distinguishability, state impurity, loss, and higher-order source emissions, all of which will degrade the quality of entangled photonic states. Partial photon distinguishability can lead to unwanted correlations between the photonic degree of freedom in which we encode our quantum information, and other internal degrees of freedom \cite{shchesnovich_partial_2015}. In heralded circuits, these correlations can result in non-computational leakage, where more than one photon occupies the modes defining a dual-rail qubit. Despite the fragility of quantum states to such noise, if physical error rates are below some threshold then quantum error correcting codes can be used to suppress logical error rates arbitrarily, permitting fault-tolerant quantum computation \cite{nielsen_fault-tolerant_2005}.  Such thresholds bound the desired fidelities of photonic resource states used as primitives in LOQC architectures. For stringent resource estimates on implementations of quantum error correcting codes in hardware, we would ideally like a detailed characterization of the noise model prescribed to our heralded physical resource states. These considerations open up the question of how small entangled photonic resources are impacted by errors, and the implications on universal photonic quantum computing architectures more broadly.

In this paper we develop a simulation framework to model entangled resource state fidelities in the presence of errors. This employs a continuous-variable (CV) representation that naturally captures realistic photonic errors before projection into the Fock space of discrete-variable (DV) entangled states. Through first quantization and representation theory, namely Schur-Weyl duality, the true nature of photonic states considering partial distinguishability and non-computational leakage errors can be revealed \cite{moylett2018,stanisic_discriminating_2018,harrow2005applications, adamson_detecting_2008, adamson_multiparticle_2007}. We benchmark entanglement generation circuits that play a key role in photonic quantum computing schemes, both in the second and first quantized state representations. Sec. \ref{IdealBellStates} details photonic state generation in the ideal case and then Sec. \ref{photonsinpractice} discusses general photon imperfections and postselected state reconstruction. The CV backend we utilise for the simulation of linear optical circuits is described in Sec. \ref{cvbackend}.
The performance of heralded linear optical entanglement generation circuits with postselection is detailed in Sec. \ref{hbsgperf}. We then analyze our resources in the first quantized picture in Sec. \ref{firstq} before finally discussing Type-II fusion gate error characterization in Sec. \ref{opint}.

\section{Ideal Bell State Generation}\label{IdealBellStates}
A common approach to encoding photonic qubits in linear optical devices is \textit{dual-rail encoding}, where two photonic modes and one photon can define a qubit. In the integrated photonics platform, qubits are often encoded over spatial modes. This allows high fidelity arbitrary single qubit operations using programmable two-port Mach-Zehnder interferometers and phase shifters. We denote the qubit basis states through physical Fock states as ${\ket{0}_L}\coloneqq{\ket{1}_1}{\ket{0}_2}$ and ${\ket{1}_L}\coloneqq{\ket{0}_1}{\ket{1}_2}$, where the occupation indicates if a photon is present in a given spatial mode (subscript). Such mode occupations where one photon is present across a dual-rail qubit, i.e., $\{\ket{0}_L,\ket{1}_L\}$, are states in the \textit{computational subspace}. There are four maximally entangled Bell states
\begin{equation}
\ket{\Psi^{\pm}}=\frac{1}{\sqrt{2}}\left(\ket{01}_L\pm{\ket{10}_L}\right),
\end{equation}
\begin{equation}
\ket{\Phi^{\pm}}=\frac{1}{\sqrt{2}}\left(\ket{00}_L\pm{\ket{11}_L}\right),
\end{equation}
with the states $\ket{\Psi^{\pm}},\ket{\Phi^{\pm}}$ all connected through single-qubit operations. We focus on \textit{heralded} states where we infer the presence of event-ready entangled states, conditioned on the detection of a subset of \textit{state herald} photons. This is in contrast to postselection, where successful gate operation is inferred by the detection of all photons in the circuit \cite{gubarev_improved_2020}. There are also hard limits on the accessible entangled photonic states in the postselected regime \cite{adcock_hard_2018}. It has been shown that heralded linear optical Bell state generation (HBSG) requires at least four photons, where two are measured in auxiliary output modes \cite{kok_linear_2007, stanisic_generating_2017}.

\begin{figure}[!t]
	\centering
	\includegraphics[width=0.95\columnwidth]{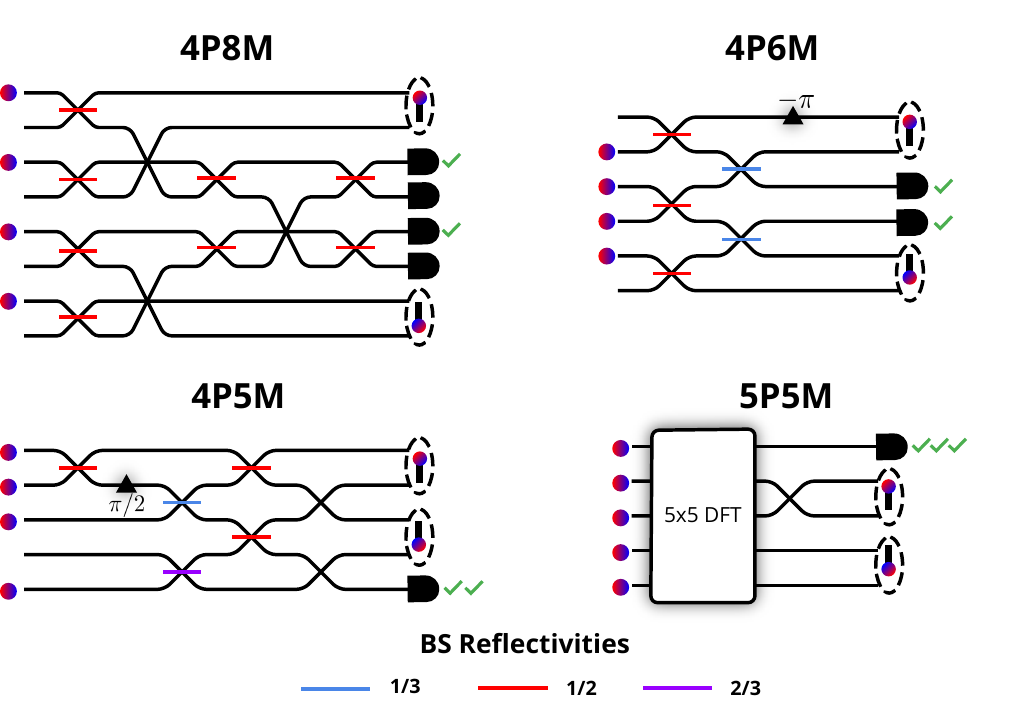}
	\makeatletter\long\def\@ifdim#1#2#3{#2}\makeatother
	\caption{\label{fig:bsgs} Heralded Bell state generation circuits which produce the $\ket{\Phi^{+}}$ Bell state when successful heralding has occurred. We denote input photons as multi-coloured to suggest potential imperfections.}
\end{figure}

There are four candidate linear optical schemes for HBSG which we model in this paper with errors. Each of these circuits takes a set of indistinguishable input photons, $\chi$, and projects a subset of these onto a Bell state. This is realised with a \textit{Positive Operator-Valued Measure} (POVM) through appropriate interferometry and measurement. The first interferometer requires four single photons and eight modes (4P8M) \cite{zhang_demonstration_2008}. Based on the measurement of two photons in two of six state herald patterns, we obtain $\ket{\Phi^{+}}$ up to permutations of the output modes (whether a permutation is required will depend on the state herald pattern). Another scheme proposed in \cite{carolan_universal_2015} also requires four input photons, this time interfered across six spatial modes, again with two photons measured (4P6M). A more recent proposal in \cite{fldzhyan_compact_2021} provides a more compact means of Bell state generation for the same number of input photons. This scheme requires a five-mode interferometer with two photons detected in a single mode (4P5M), exhibiting four-photon interference, unlike in the 4P8M and 4P6M schemes which only contain two and three-photon interference respectively. In \cite{fldzhyan_compact_2021} it is shown that the 4P6M scheme can be partitioned into two three-photon interference schemes, whereas the 4P5M scheme cannot be partitioned further. The final Bell state scheme we analyze is based on a proposal for arbitrary dimensional heralded GHZ state generation, exploiting the zero transmission law in discrete Fourier transforms (DFT) \cite{paesani_scheme_2021}. Five photons interfere in a $5\times5$ DFT unitary (5P5M), and the detection of three photons in a single mode leaves the remaining photons in a Bell state. Each of these schemes is shown in Fig.\ref{fig:bsgs}. The success probabilities of these HBSGs (for a fixed heralding pattern) are approximately $0.031, 0.074, 0.11$ and $0.096$ respectively \cite{successprobkey, losimgit}. Whilst the success probabilities of these schemes are not equal and lead to different resource requirements for near-deterministic generation with multiplexing, here we assume that we have a set of detection events ideally leading to heralding of our desired Bell state, $\ket{\Phi^{+}}$. In Appendix \ref{ghzapp} we discuss 3-(qubit)GHZ state generation (6P12M) which can be seen as a natural extension of the 4P8M HBSG scheme.

\section{Photons in practice}\label{photonsinpractice}
Indistinguishable single photons are a critical resource for photonic quantum information processing. If photons are partially distinguishable in internal degrees of freedom such as spectrum, which-path information will degrade subsequent quantum interference between these photons. Here we introduce models for partially distinguishable photons and describe how this is manifested in realistic photon sources. We then discuss source-agnostic state reconstruction techniques which are usually applied in LOQC.
\subsection{Photon distinguishability and sources}\label{distandsources}
Two photonic distinguishability structures that have been proposed are the \textit{orthogonal bad-bits} (OBB) and \textit{random source} (RS) models \cite{sparrow_quantum_2017}. In the OBB model each photon's state is  $a_{\gamma}^{\dag}[\psi_i]\ket{0}$, with $a_{\gamma}^{\dag}[\psi_i]$ describing a creation operator for path $\gamma$ where the $i^{th}$ photon's pure internal state is
\begin{equation}\label{obbeq}
\ket{\psi_i}=\mathcal{V}^{1/4}\ket{\xi_0}+(1-\sqrt{\mathcal{V}})^{1/2}\ket{\xi_i},
\end{equation} consisting of a common state $\ket{\xi_0}$ and a contribution from an orthonormal state $\ket{\xi_i}( \braket{\xi_i}{\xi_j}=\delta_{ij})$. In this model the HOM visibility for a photon pair is given as $\mathcal{V}_{ij}=\lvert{{{\braket{\psi_i}{\psi_j}}}}\rvert^2=\mathcal{V}, i \neq j$.
In the RS model, photon sources are assumed to produce independent and identically distributed (i.i.d) photons in internal states $\nu(\vec{x})$ with probability $P(\vec{x})$, where $\vec{x}$ are variables that characterize the photon's internal state. By assuming independent sources we can define a common average state for each photon through
\begin{equation}\label{rseq}
\rho=\sum_{\Vec{x}}\text{P}(\Vec{x})\nu(\Vec{x})=\sum_{i} p_i\ket{\xi_i}\bra{\xi_i}.
\end{equation}
Here, the average visibility for photons from any pair of sources is given by $
\mathcal{V}=\Tr(\rho^2)=\sum_i{p_i^2}$.
For the OBB model, as the internal state contribution for each photon comes from an orthonormal basis, they do not interfere with one another. This leads to the OBB and RS models being equivalent in the high visibility regime where for the OBB model, $\mathcal{V}\sim 1$, and for the RS model, $p_{0} \gg p_{i\neq{0}}$, since then the most likely error is to have a single photon with an internal state that is distinguishable from the others \cite{sparrow_quantum_2017}.

These models of imperfect photons can be associated more explicitly to specific photon sources. Common approaches for photon generation use material non-linearities, producing photon pairs where a signal photon is heralded to infer the presence of an event-ready idler photon (\textit{source heralding}). For HBSG circuits each photon in $\chi$ will require the detection of a signal photon to ensure the correct number of photons are injected into the circuit. Examples of spontaneous photon generation processes are parametric down conversion (SPDC) and four-wave mixing (SFWM), with interaction Hamiltionians that represent the generator of multimode two-mode squeezing operators as
\begin{equation}\label{intham}
\hat{H}=\iint{d\nu_{1}d\nu_{2}}F(\nu_1,\nu_2)a_{1}^{\dag}(\nu_1)a_{2}^{\dag}(\nu_2)+h.c,
\end{equation}
with $F(\nu_1,\nu_2)$, the photon-pair correlation function known as the joint-spectral amplitude (JSA). This function can encapsulate the quality of photon sources based on two-mode squeezers \cite{thomas_general_2020} with the Schmidt decomposition \cite{grice_spectral_1997,law_continuous_2000} of the JSA given by 
\begin{equation}\label{jsaeq}
    F(\nu_1,\nu_2)=\sum_{k}{s_k}{\psi_k}(\nu_1){\phi_k}^{*}(\nu_2),
\end{equation}
where $\{s_k\}$ are positive Schmidt coefficients and $\{{\psi_k}\},\{{\phi_k}\}$ are sets of orthonormal functions. Describing the biphoton state $\ket{\zeta}$ (low-squeezing regime) through evolution of the vacuum (by exponentiation of the interaction Hamiltonian) with broadband mode operators $\mathcal{A}_{k}^{\dag}=\int{d\nu_{1}{\psi_k}(\nu_1)a_{1}^{\dag}(\nu_1)}$ and $\mathcal{B}_{k}^{\dag}=\int{d\nu_{2}{\phi_{k}^{*}}(\nu_2)a_{2}^{\dag}(\nu_2)}$ gives
\begin{equation}\label{biphoton}
\begin{gathered}
\ket{\zeta}=e^{-i\hat{H}}\ket{vac}\\\approx\ket{vac}-iF(\nu_1,\nu_2)\iint{d\nu_{1}}{d\nu_{2}}{a_{1}^{\dag}(\nu_1)}{a_{2}^{\dag}(\nu_2)}\ket{vac}.
\end{gathered}
\end{equation}
Multiple Schmidt terms lead to an entangled biphoton state, projecting the idler into a spectrally mixed state $\rho_i$, on detection of the signal photon \cite{thomas_general_2020}. Discretizing Eq. \ref{intham} provides a singular value decomposition (SVD) of Eq. \ref{jsaeq} for the multimode two-mode squeezing operators in terms of independent two-mode squeezers acting on orthogonal Schmidt modes. From the above discussion the coefficients $\{p_i\}$ for RS photons are equivalent to $\{s_k^2\}$. The purity $\mathcal{P}$ of $\rho_{i}$ is given as $\Tr({\rho_i}^2)=\sum_k{{s_k}}^4$ \cite{graffitti_design_2018}.

The other prominent approach to generating photons is exploiting deterministic emitters. Solid-state systems such as quantum dots (QDs) have opened the path towards high quality single photon sources due to their high single photon generation rates and true single photon emission \cite{hoang_ultrafast_2016, yamamoto_single_2005}. Phonon-mediated dephasing can lead to correlations between the phonon and single emitted photon energies leading to spectral mixedness, analogous to the state structure in Eq. \ref{biphoton} for spontaneous sources \cite{tiurev_fidelity_2021}. For single photons emitted from separate QDs, imperfections due to differing emission frequencies lead to coherent distinguishability in their internal states analogous to the OBB model. 

Photon quality from both spontaneous and deterministic sources can be improved through a variety of techniques. Methods include spectral or temporal filtering \cite{massaro_improving_2019,legero2003time}, at the cost of reduced count rates, distillation techniques \cite{sparrow_quantum_2017,marshall_distillation_2022}, which require many noisy single photon states, and, for coherent distinguishability (OBB), high temporal resolution detection \cite{yard2022onchip,zhao2014entangling}.


\subsection{Tomography of photonic qubits}\label{statetomo}
Reconstructing an unknown quantum state $\rho$ is known as \textit{quantum state tomography}, allowing for the benchmarking of quantum information experiments \cite{dariano_quantum_2003}. As we can never perfectly distinguish between non-orthogonal states such as $\ket{0}$ and $\ket{+}$ with a single measurement, state tomography relies on preparing multiple copies of $\rho$ and expanding it in a complete Hermitian operator basis for the state space. In the qubit case, this operator basis is the Pauli group, $G=\{\mathds{1},\sigma_x,\sigma_y,\sigma_z\}$ where $\mathds{1}$ is the $2 \times 2$ identity matrix and $\sigma_x,\sigma_y,\sigma_z$ are the $2 \times 2$ Pauli matrices. The density matrix, $\rho_{n,2}$ of $n$ qubits can be defined as
\begin{equation}
    \rho_{n,2}=\frac{1}{2^n}\sum_{i_{1},\ldots,i_{n}=0}^{3}r_{i_{1},\ldots,i_{n}}\,\mathcal{G}_{i_1}\otimes\ldots\otimes{\mathcal{G}_{i_n}}.
\end{equation}
Full tomographic reconstruction requires determining the expectation values of the observables
\begin{equation}\label{expvalstomography}r_{i_{1},\ldots,i_{n}}=\langle{\mathcal{G}_{i_{1}}}\otimes...\otimes{\mathcal{G}_{i_{n}}}\rangle.
\end{equation}
This requires an exponential number of measurements and so quickly becomes experimentally unfeasible \cite{thew_qudit_2002}. For many states that occur in practical situations however, only a small number of parameters is required to fully describe the state. This leads to more efficient approaches for tomographic reconstruction such as for matrix product states \cite{cramer_efficient_2010}. Low entropy (pure or nearly pure) quantum states may be reconstructed via tomographic techniques based on compressed sensing \cite{gross_quantum_2010}. Pure quantum states may also be described through the set of operators they are \textit{stabilized} by (a $+1$ eigenstate of), with these states forming the basis of many quantum error correcting codes \cite{gottesman_stabilizer_1997}.

\section{The CV Backend}\label{cvbackend}
The assessment of imperfect heralded resource states in linear optical circuits requires a robust simulation framework. Following the approach of Quesada et al., we conditionally prepare heralded non-Gaussian states with squeezed Gaussian input states in the CV regime \cite{quesada_simulating_2019}. The non-Gaussianity we refer to here is that of Fock states which can be projected out from Gaussian states through the POVM generated by interferometry and heralding. This framework can accommodate both the CV nature of spontaneous squeezed sources based on nonlinear frequency conversion and also the DV states generated from single emitters (by using lossless photon-number-resolving detectors (PNRDs)). Here we describe the simulation methodology and backend design considerations required to reconstruct non-Gaussian states.

\subsection{Non-Gaussian state preparation}
Heralded non-Gaussian states can be conditionally prepared from a Gaussian state defined in terms of its \textit{displacement vector} and \textit{covariance matrix} $({\Vec{\beta}},\sigma)$ respectively and PNRDs as described in \cite{sabapathy_production_2019}. This requires computing an expression for the Fock basis matrix elements $\matrixel{\textbf{m}}{\hat{\varrho}}{\textbf{n}}$ with $\textbf{m},\textbf{n}$ representing photon occupations in each of the $l$ modes comprising the Gaussian state $\hat{\varrho}$. Such elements can be calculated through
\begin{equation}\label{matrixelem}
\matrixel{\textbf{m}}{\hat{\varrho}}{\textbf{n}}=T \times \text{lhaf}(\Tilde{A}).
\end{equation}
The quantities $T, \Tilde{A}$, and the loop hafnian (lhaf) matrix function that counts perfect matchings in weighted graphs with loops are detailed in  Appendix \ref{cv-app}. The number of steps to compute Eq. \ref{matrixelem} is exponential in $l$. Resource states that we will be assessing in Sec. \ref{hbsgperf} onwards will be defined over $k \subset l$ modes, based on the detection and heralding of photons in modes $\textbf{n}_{h}=(\textbf{n}_1,\textbf{n}_2,...,\textbf{n}_{l-k})$. By the partial measurement of $\hat{\varrho}$ on $\textbf{n}_{h}$, we can reconstruct Fock basis matrix elements over the heralded non-Gaussian state $k$-mode subspace.

\begin{figure*}
	\centering
	\includegraphics[width=0.95\textwidth]{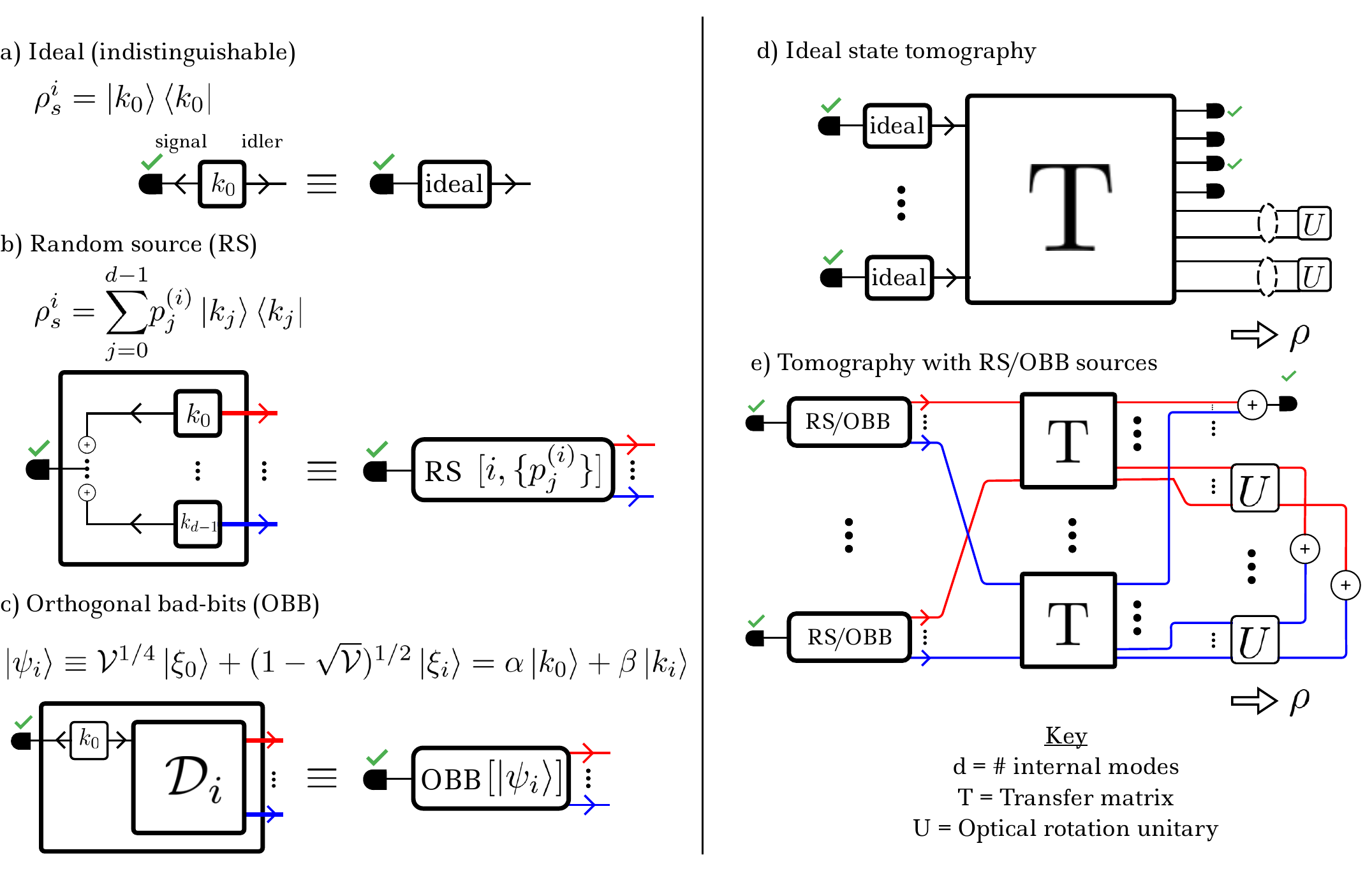}
	\makeatletter\long\def\@ifdim#1#2#3{#2}\makeatother
	\caption{\label{fig:RSOBBfig}\textbf{a-c)} Photon sources in the ideal, RS and OBB cases which are used in the CV backend for state reconstruction. \textbf{d)} The ideal reconstruction requires $|\chi|$ photon sources and state reconstruction is performed via tomography. \textbf{e)} Reconstructing the logical state from imperfect sources is performed by dilating the interferometer to account for the internal photon state information before incoherently summing logically equivalent contributions.}
\end{figure*}

\subsection{Modelling imperfect sources}
\subsubsection{Random source}\label{rsmodellingsec}
To map the RS photon purity to the CV backend we note that the Schmidt coefficients $\{{s_k}\}$ can be represented by the probability of generating a photon pair in a given Schmidt mode as
\begin{equation}\label{rspurityeq}
\mathcal{P}=\sum_{i=0}^{d-1}{q_{i}}^2,
\end{equation}
with $q_{i}$ defined as the normalised probability that a single photon pair is generated from a squeezed state, $p_{i}$, in the $i^{th}$ Schmidt mode as
\begin{equation}
q_{i}=\frac{p_i}{\sum_{i=0}^{d-1}{p_i}},
\end{equation}
\begin{equation}
p_{i}={\kappa_i}^2(1-{\kappa_i}^2),
\end{equation}
where $\{p_i\}=\{s_k^2\}$ and $\kappa=\tanh{r}\in[\,0,1]\,$ is the parameter quantifying the two-mode squeezing. The set of indistinguishable photons, $\chi$, required for the $m$-mode entanglement generation circuits in Sec. \ref{IdealBellStates} can now be modelled, with arbitrary purity $\mathcal{P}$, through i.i.d distinguishable photon sources, with non-zero photon generation probabilities across each Schmidt mode. For each photon source required, we herald on a signal photon. We then simulate the RS mixed idler photon by running $d$ copies of the ideal circuit in parallel and incoherently summing the results (see Fig. \ref{fig:RSOBBfig}b,e). For indistinguishable photons we have one Schmidt mode (as seen in Fig. \ref{fig:RSOBBfig}a,d) so only require one copy of the circuit.

\subsubsection{Orthogonal bad-bits} 
In the OBB model, the photon described through $a_{\gamma}^{\dag}[\psi_i]\ket{0}$ is prepared in the pure state $\ket{\psi_{i}}$, comprising a component shared with all other photons $\ket{\xi_{0}}$ and a unique distinguishable state $\ket{\xi_{i}}$ (see Eq. \ref{obbeq}). The different distinguishable states can be associated with separate modes the photons can occupy. The states $\ket{\psi_{i}}$ can be prepared by injecting an indistinguishable photon into a \textit{distinguishability unitary} $\mathcal{D}_i$ that spreads the photon across these orthogonal modes defined as
\begin{equation}
\mathcal{D}_i=
\begin{bmatrix}
\alpha  & \cdots  & \beta & & & \\
\vdots  &  \ddots & \vdots & & & \\
\beta  & \cdots  & -\alpha & & & \\
& & & 1 & & \\
& & & & \ddots & \\
& & & & & 1\\
\end{bmatrix},
\end{equation}
where $\alpha,\beta$ represent beamsplitter matrix elements implemented between the $0^{th}$ and $i^{th}$ modes (see Fig. \ref{fig:RSOBBfig}c). Once the distinguishability has been set, the interferometer is applied to the spatial modes for a given distinguishable state. Similar to Schmidt contributions in the RS model, these evolutions occur independently within an $m({|\chi|}+1){\times}m({|\chi|}+1)$ interferometer (excluding source heralds), where $({|\chi|}+1)$ is the dimension required to faithfully model an OBB photonic internal state structure with a complete orthonormal basis.

\begin{figure*}
	\centering
	\includegraphics[width=0.95\textwidth]{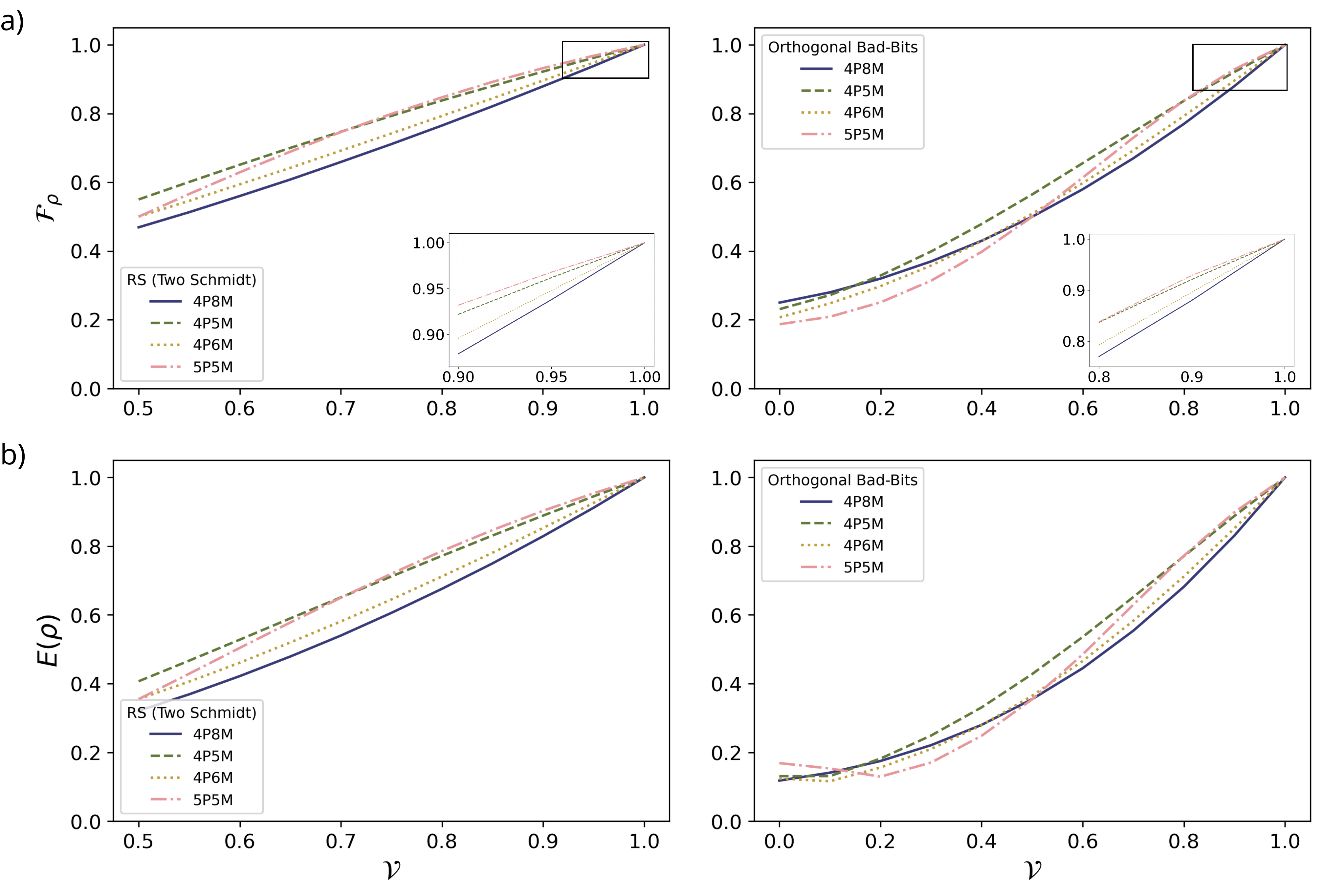}
	\caption{\label{fig:2ndQRSOBB}Performance benchmarking of the HBSG circuits for the two proposed distinguishability models -- RS and OBB. \\ \textbf{a)} Logical Bell state fidelities $\mathcal{F}_{\rho}$ as a function of the pairwise photon visibility, $\mathcal{V}$. \textbf{b)} Entanglement of formation $E(\rho)$ over the same visibility range.}
\end{figure*}

\subsection{State reconstruction}\label{staterecon}
Tomography provides a computational state benchmark from which to assess the impact of distinguishability on logical state fidelities. Details of the CV tomographic reconstruction are provided in Appendix \ref{cvtomo}, with optical rotation unitaries appended to modes defining logical qubits (as shown in Fig. \ref{fig:RSOBBfig}d,e).  Logically equivalent internal state contributions to the same qubit basis are incoherently summed to reconstruct the state in the computational subspace in a postselected manner. When performing tomography we require independent simulations of the linear optical circuit for the logically equivalent source herald patterns, known as \textit{source firings}. Naively we would simulate all state herald patterns conditioned on the set of source firings but we filter these as described in Appendix \ref{sourcefiringlogic} to provide a reduction in the time overhead for state reconstruction using Eq. \ref{matrixelem}.

\begin{figure*}
	\centering
	\includegraphics[width=0.95\textwidth]{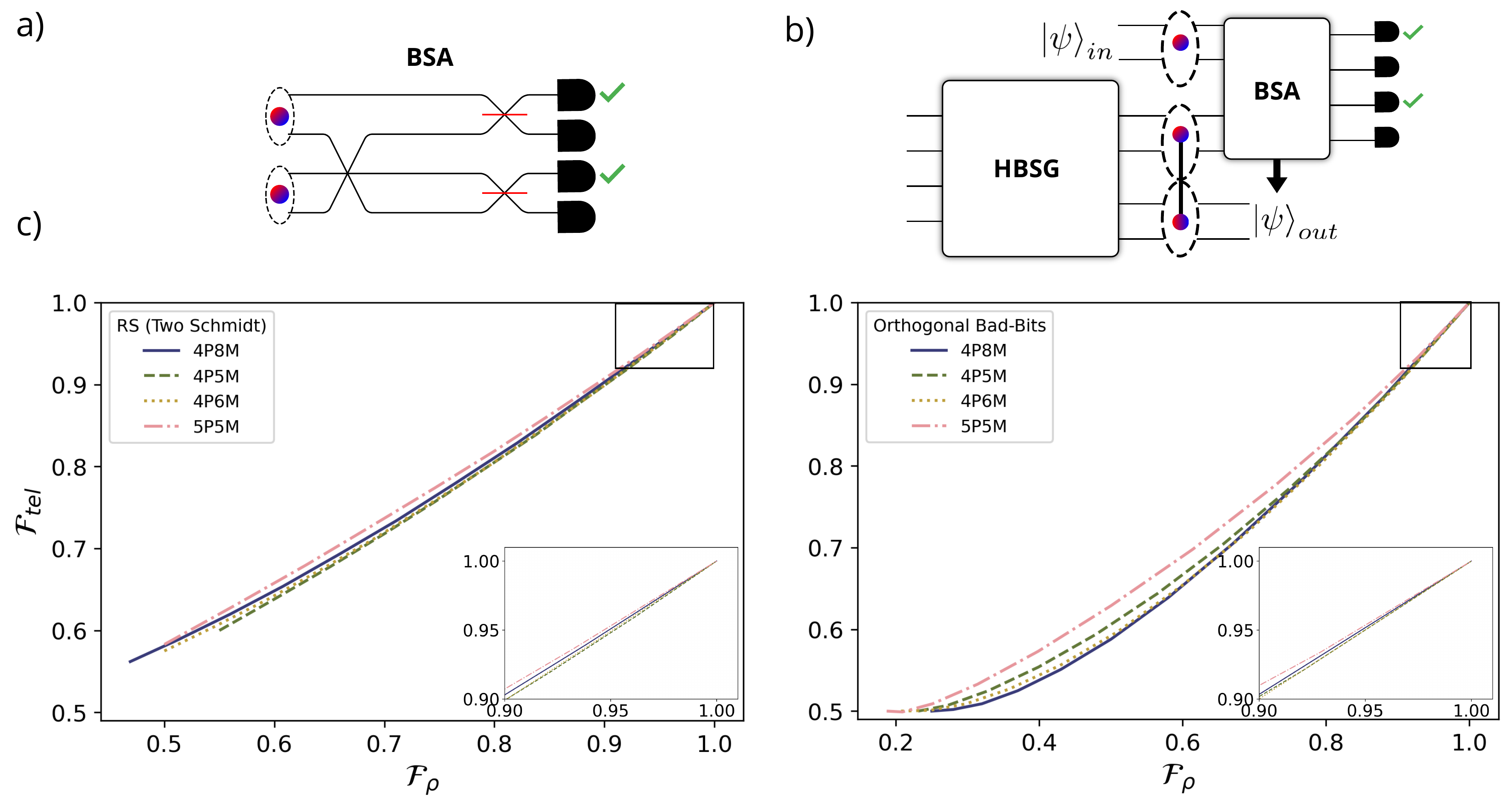}
	\caption{\label{fig:telep}\textbf{a)} Linear optical BSA for discriminating Bell states which operates with a success probability of $0.5$. \textbf{b)} Optical implementation of the teleportation of an arbitrary single photonic qubit. The general HBSG utilized here can be represented by any of the proposed HBSGs. The output measurements of the BSA permit teleportation without prior knowledge of the state. \textbf{c)} Teleported single qubit fidelity $\mathcal{F}_{tel}$ of the $\ket{+}$ state as a function of $\mathcal{F}_{\rho}$ for both RS (two Schmidt) and OBB photon sources ($\rho$ used as the entangled resource in the protocol).}
\end{figure*}

\section{Performance benchmarking of circuits with state tomography}\label{hbsgperf}
The CV backend provides us with a framework to model the effects of imperfections in heralded entanglement generation circuits. Here we assess the performance of proposed HBSGs and optical quantum teleportation in the presence of partial distinguishability errors. In Appendix \ref{threshandloss} we discuss photon loss and multi-photon contamination in the presence of threshold detection.

\subsection{Resource generation}\label{resourcegen}
Simulating each HBSG introduced in Sec. \ref{IdealBellStates} for the RS (two Schmidt) and OBB models, in Fig. \ref{fig:2ndQRSOBB}a we report the logical state degradation for decreasing photon visibility, $\mathcal{V}$ (increasing distinguishability). The physically accessible visibility range for the RS (two Schmidt) model is $0.5 \leq V \leq 1$ with $\mathcal{V}=0.5$ equivalent to two equal Schmidt terms (maximal spectral mixedness). The minimum $\mathcal{V}$ for $K$ Schmidt modes is $1/K$. OBB photons are modelled for $0 \leq V \leq 1$ where for $\mathcal{V}=0$ each photon is defined completely in terms of the distinguishable orthonormal contribution $\ket{\xi_i}$. The choice of $K=2$ for the RS simulation enables tractable simulations on a standard laptop.  In Appendix \ref{34SchmidtApp} we report results for the RS model with a filtered waveguide source requiring $K > 2$ Schmidt modes. Whilst adding additional Schmidt modes slows down state reconstruction, the CV backend can simply account for additional Schmidt contributions. Each data point comes from a complete simulation of the interferometer for a given visibility over all permitted source firings. We compute the computational fidelity,  $\mathcal{F}_{\rho}$, based on the state overlap with ${\ket{\Phi^{+}}}$ as $\mathcal{F}_{\rho}=\expval{\rho}{\Phi^{+}}$ with $\rho$ obtained via standard tomography as described in Sec. \ref{staterecon} and depicted in Fig. \ref{fig:RSOBBfig}e. For indistinguishable photons with $\mathcal{V}=1$, as expected we obtain unity fidelities for both the RS and OBB models. In Appendix \ref{rsandobbapp} we show that $\mathcal{F}_{\rho}$ values for RS and OBB sources are approximately equal for each HBSG for the visibility regime where both source models are physically accessible.

In the high visibility regime (see insets Fig. \ref{fig:2ndQRSOBB}a), the 5P5M scheme shows the least logical degradation, whereas the 4P8M scheme is the most susceptible to source photon imperfections. As we move to the highest photon-number interference exhibited in any of the HBSG schemes, then we observe increased robustness to partial distinguishability. At lower visibilities, whilst for the RS model similar relative performance between circuits is maintained, the 4P8M scheme is degraded the least by OBB partial photon distinguishability. For completeness, we model over the full visibility regime; however, the main region of interest is the high visibility regime where candidate sources for scalable photonic quantum computing should operate.
Another metric for assessing the robustness of the HBSGs to photon imperfections is the \textit{entanglement of formation} (EoF), an entanglement monotone for bipartite quantum states \cite{wootters_entanglement_1998}. The EoF asymptotically quantifies the number of Bell states needed to prepare $\rho$ using local quantum
operations and classical communication \cite{bennett_mixed-state_1996}. As $\rho$ can be realized through a number of pure state decompositions, the EoF quantifies the average entanglement of the pure state decomposition minimized over decompositions of $\rho$ as
\begin{equation}\label{eofeq}
E(\rho)=\min\sum_{i}{p_i}E(\psi_i).
\end{equation}
In Fig. \ref{fig:2ndQRSOBB}b, the EoF is computed for each circuit, displaying the same relative robustness as the logical state fidelities for both RS and OBB photons. $E(\rho)$ is monotonic in the \textit{concurrence} entanglement measure \cite{nielsenandchuang}.

\subsection{Optical quantum teleportation}\label{opticalquantumtelep}
The transfer of a quantum state $\ket{\psi}$ from one location to another disparate location is known as \textit{quantum teleportation}. Teleportation can be performed without the owner of  $\ket{\psi}$ having any prior knowledge of the state. The two parties, Alice and Bob, each hold half of a qubit Bell state \cite{bennett_teleporting_1993}. The entire system is in a pure product state $\ket{\psi}\otimes\ket{\Phi^+}$, with no classical correlations or entanglement between these two states. Alice then performs a Bell state measurement (BSM) on her half of the entangled state ${\ket{\Phi^{+}}}$ and the unknown state $\ket\psi$ she wants to teleport. Irrespective of the measurement outcome, a classical communication channel provides Bob with information to apply the required rotation for perfect reconstruction of $\ket{\psi}$ on his qubit. By changing the basis of her BSM, Alice can implement a unitary rotation on $\ket{\psi}$, with this principle forming the basis of computation in MBQC. Optically, we can perform teleportation with a linear optical Bell state analyzer (BSA) representing Alice's measurement device and a shared Bell state ${\ket{\Phi^{+}}}$ produced by a HBSG. In photonics, it is impossible to perfectly discriminate Bell states. The optimal, ancilla-free, probabilistic BSA is shown in Fig. \ref{fig:telep}a, and operates with a success probability of $0.5$ \cite{calsamiglia_maximum_2001, braunstein_measurement_1995, weinfurter_experimental_1994}.

In Fig. \ref{fig:telep}b we show the optical circuit for teleportation of a single arbitrary photonic qubit, utilising a general HBSG as the entangled state resource generator. If we assume each photon required for teleportation is produced with an imperfect source, then we can compute the fidelity $\mathcal{F}_{tel}$ of a teleported $\ket{+}$ state for each HBSG circuit. In Fig. \ref{fig:telep}c we model the performance of each HBSG for optical quantum teleportation with RS and OBB photon sources in terms of $\mathcal{F}_{\rho}$, assuming correct heralding on the output of the BSA. As all HBSGs perform comparably for high fidelity resource states (see Fig. \ref{fig:telep}c insets), the choice of which generator to use for a specific teleportation application requires optimization of other parameters such as circuit success probability or optical depth, relative to the specific hardware used. For example, whilst minimising the number of photons and modes leads to the 4P5M HBSG as the desirable candidate, there are trade-offs such as in heralding on more than one photon in a single mode that must be considered. In terms of input photon visibility, relative HBSG robustness reflects that of the state generation circuits. In Appendix \ref{supp-telep-info} we model the robustness of each of the sets of antipodal single qubit eigenstates of the Pauli matrices utilising the 4P8M HBSG as the entangled resource generator.

\section{Distinguishability in first quantization}\label{firstq}
The discussion so far has been in the second quantized formalism of photonic creation and annihilation operators. Now we view bosonic Fock states in first quantization where the impact of particle distinguishability and symmetries is more explicit. In this section we motivate the limitations of standard tomographic reconstruction following the discussions by Adamson et al.~\cite{adamson_detecting_2008, adamson_multiparticle_2007}. We then formalize the mapping between second and first quantized partially distinguishable photonic states following \cite{stanisic_discriminating_2018}. Once these details are laid out we describe how this is implemented in the CV backend, making use of efficient Schur-Weyl transforms \cite{bacon_efficient_2006, bacon_quantum_2005}, thus revealing the impact on resource state fidelities.
\subsection{Limitations in tomographic reconstruction}
Quantum state tomography of photons is typically performed in a postselected manner with the detection apparatus constraining the reconstructed state to lie in the multi-qubit computational subspace. However, in the presence of errors, the carefully engineered interference in state generation circuits degrades, allowing the photonic state to leak into the full bosonic space that contains non-computational terms. The characterization of such states was proposed and carried out (for two-photon polarization-encoded states) in \cite{adamson_multiparticle_2007, adamson_detecting_2008}. Here a tomographic method is developed that faithfully accounts for state degradation from both decoherence and internal state distinguishability, treating each as distinct phenomena, in the first quantized paradigm. We leverage this state reconstruction technique and analyze the impact on entanglement generation circuits. Utilising this photonic state description, we aim to better understand the impact of errors in heralded circuits generating resources used in architectures for universal LOQC, than with standard postselected tomographic approaches to state reconstruction.
\begin{figure}[!t]
	\centering
	\includegraphics[width=0.475\textwidth]{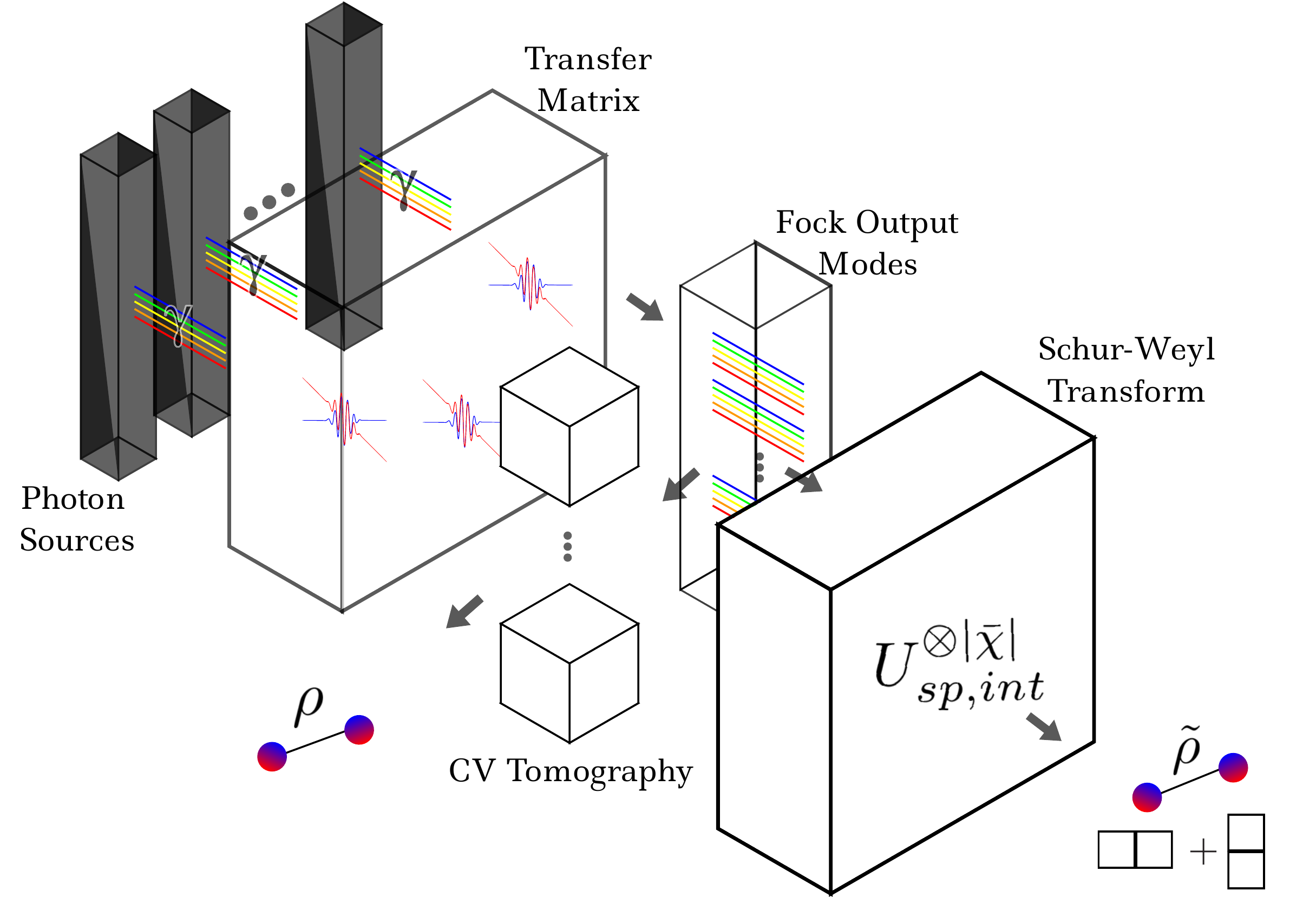}
	\caption{\label{fig:conceptfig}Decomposition of the CV backend for simulating heralded linear optical circuits in the presence of imperfections. The unitary Schur-Weyl transform, $U_{sp,int}^{\otimes{\abs{\bar{\chi}}}}$ is performed on the Fock space representation at the output of the transfer matrix to benchmark logical state degradation relative to the CV tomography method.}
\end{figure}
\begin{figure*}
	\centering
	\includegraphics[width=0.95\textwidth]{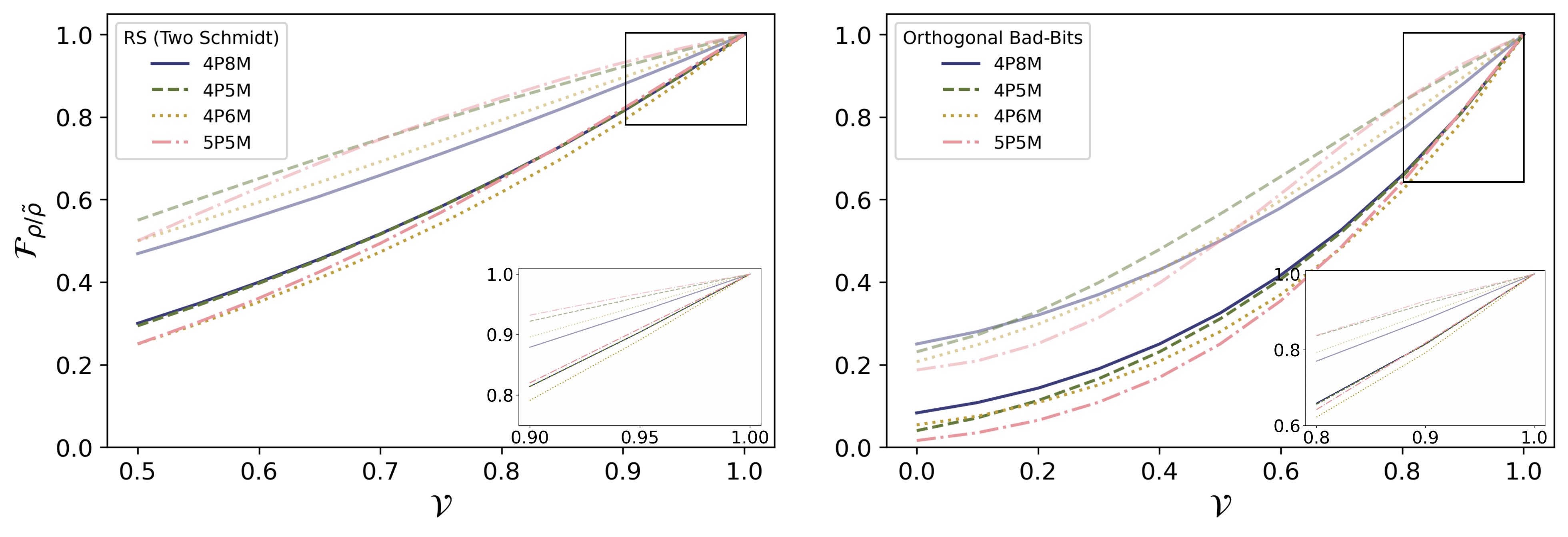}
	\caption{\label{fig:1stQRSOBB}Logical state fidelities $\mathcal{F}_{\tilde{\rho}}$ for HBSG circuits for the RS and OBB source models. First quantized logical states $\tilde{\rho}$ are reconstructed using the CV backend and QuTip library, benchmarked relative to states reconstructed from standard tomography $\rho$ (we also plot $\mathcal{F}_{\rho}$ for comparison).}
\end{figure*}

\subsection{Photonic states - from second to first quantization}
The mapping from second to first quantized photonic states requires taking the standard mode occupation description of second quantized Fock states and expressing this explicitly in terms of particle wavefunctions. Let us consider a distinguishable state of two photons in different spatial (system) modes ($1,2$) with orthogonal internal state (label) modes ($A,B$). Here we assume that the system is the degree of freedom resolved upon detection, whereas the label is not. This state can be described by creation operators as  ${a_{1,A}^{\dag}}{a_{2,B}^{\dag}}\ket{0}$. Obeying bosonic symmetries, this state is mapped to a symmetrised two-particle state.  We can explicitly construct the state and separate the system and label degrees of freedom \cite{stanisic_discriminating_2018} as
\begin{equation} \label{distpoleq}
\begin{gathered}
\ket{\psi_{12}^{AB}}=\ket{1,A}\vee\ket{2,B}, \\
=\frac{1}{\sqrt{2}}(\ket{1,A}\otimes\ket{2,B}+\ket{2,B}\otimes\ket{1,A}),
\end{gathered}
\end{equation}
where $\vee$ is the symmetric tensor product and each state ket corresponds to a single particle with Hilbert space $\mathcal{H}^{(1)}=\mathcal{H}_S\otimes\mathcal{H}_L$. This state notation allows us to see how symmetry is distributed across each degree of freedom. Separating the resolved and unresolved degrees of freedom for the two-particle system-label state gives
\begin{equation}
    \ket{\psi_{12}^{AB}}=\frac{1}{\sqrt{2}}(\ket{\psi_{12}^{+}}\otimes\ket{\psi_{AB}^{+}}+\ket{\psi_{12}^{-}}\otimes\ket{\psi_{AB}^{-}}),
\end{equation}
where the two-particle state for a single degree of freedom is $\ket{\psi_{ij}^{\pm}}=(\ket{i,j}\pm\ket{j,i})/\sqrt{2}$ with $(-)+$ corresponding to (anti-)symmetric particle states respectively. We observe that whilst the state is still symmetric overall, it has components in the anti-symmetric parts of the two-particle joint Hilbert space. Note, if we were to apply the same mapping procedure to the indistinguishable ${a_{1,A}^{\dag}}{a_{2,A}^{\dag}}\ket{0}$ state then this only occupies the symmetric two-particle subspace. The extension of the state outside of the symmetric Hilbert space is only seen when the labels on each particle are different -- distinguishability has led to system-label space correlations \cite{stanisic_discriminating_2018, turner_postselective_2016}.


Assessing the impact of photon distinguishability will require the subsequent unitary evolution of symmetric states composing the system and label. \textit{Schur-Weyl duality} \cite{rowe_dual_2012, bacon_quantum_2005} permits separate decompositions of the system and label degrees of freedom in the total $N$-particle Hilbert space $\mathcal{H}_{SL}^{(N)}$ as
\begin{equation}\label{hilbertdecomp}
    \begin{gathered}
    \mathcal{H}_{SL}^{(N)}=(\mathcal{H}_S)^{\otimes{N}}\otimes(\mathcal{H}_L)^{\otimes{N}}\\
    =(\underset{\lambda}{\oplus}\mathcal{H}_{S}^{\lambda})\otimes(\underset{\lambda'}{\oplus}\mathcal{H}_{L}^{\lambda'}).
    \end{gathered}
\end{equation}

The irreps $\lambda,\lambda^{\prime}$, describing subspaces of $\mathcal{H}_{SL}^{(N)}$ which transform amongst themselves are reviewed in Appendix \ref{SchurWeyl}, along with the construction of symmetrised $N$-particle states and Schur-Weyl duality more broadly. Evolution of symmetric system-label states through ${U}_{SL}$ decomposes in the same fashion as the $N$-particle Hilbert space,
\begin{equation}
U_{SL}^{\otimes{N}}\simeq\underset{\lambda}{\oplus}(U_{S}^{\lambda}\otimes{U}_{L}^{\lambda}).
\end{equation}
Here, $\simeq$ suggests equality up to a basis transformation. Known as \textit{unitary-unitary duality} \cite{rowe_dual_2012}, unitary actions in the symmetric multi-particle space will act on irreps independently. This unitary takes a block-diagonal form in the Schur-Weyl basis with each block corresponding to a unique irrep \cite{stanisic_discriminating_2018}. For  $U_{SL}^{\otimes{2}}$, matrix elements in the symmetric and anti-symmetric subspaces are transformed through permanents and determinants of the unitary transformation respectively \cite{turner_postselective_2016}. Obtaining the reduced system state through tracing out the label can now leave the state with population outside of the symmetric irrep of the system subspace. Through a simple description of the photonic input state and interferometer in the presence of physical errors, we now consider non-zero occupation across all accessible irreps. Furthermore, in comparison to having to construct the transfer matrix over the whole Hilbert space, we require the computation of a set of smaller matrices  \cite{stanisic_2020}.

\subsection{Schur state fidelities}
Photons defining our heralded resource state can be denoted with $\bar{\chi} = \chi\setminus{h}$, where $h$ is the set of photons required for state heralding on the set of input photons to the circuit, $\chi$. We can map our system of $\abs{\bar{\chi}}$-photon states with internal dimension $d$ to symmetric $\abs{\bar{\chi}}$-particle states, defined over the $\mathcal{H}_{SL}^{(\abs{\bar{\chi}})}$ space through the Schur-Weyl transform. For partially distinguishable photonic states we define the system as the spatial degree of freedom, $sp$, and label as the internal degree of freedom, $int$. As the CV backend projects out $h$, we only require the Schur-Weyl transform on $\bar{\chi}$ before tracing out the internal degree of freedom. The unitary evolution of the symmetric spatial-internal photonic output state will be given by
\begin{equation}
U_{sp,int}^{\otimes{\abs{\bar{\chi}}}}\simeq\underset{\lambda}{\oplus}(U_{sp}^{\lambda}\otimes{U}_{int}^{\lambda}),
 \end{equation}
where ${U}_{int}^{\lambda}=\mathds{1}$ if we assume interferometry is independent of the internal states. The Schur-Weyl transform is shown as part of the CV backend in Fig. \ref{fig:conceptfig}. To perform this transformation we require the full Fock representation at the output, capturing both photon terms in the computational subspace and non-computational terms (where both photons in $\bar{\chi}$ occupy a single dual-rail qubit). The implementation of Schur-Weyl transforms will be given by fixed unitary transformations for a given $\abs{\bar{\chi}}, d$ which can be stored for later use \cite{bacon_efficient_2006}. We use the \textit{QuTip} Python library \cite{johansson_qutip_2012} to perform the mapping  and in Appendix \ref{SchurPseudoApp} we provide a pseudocode example of the Schur-Weyl transform to show how we reconstruct reduced logical Schur states, which we generally refer to hereafter as $\tilde{\rho}$.

\begin{figure}[!t]
	\centering
	\includegraphics[width=0.74\columnwidth]{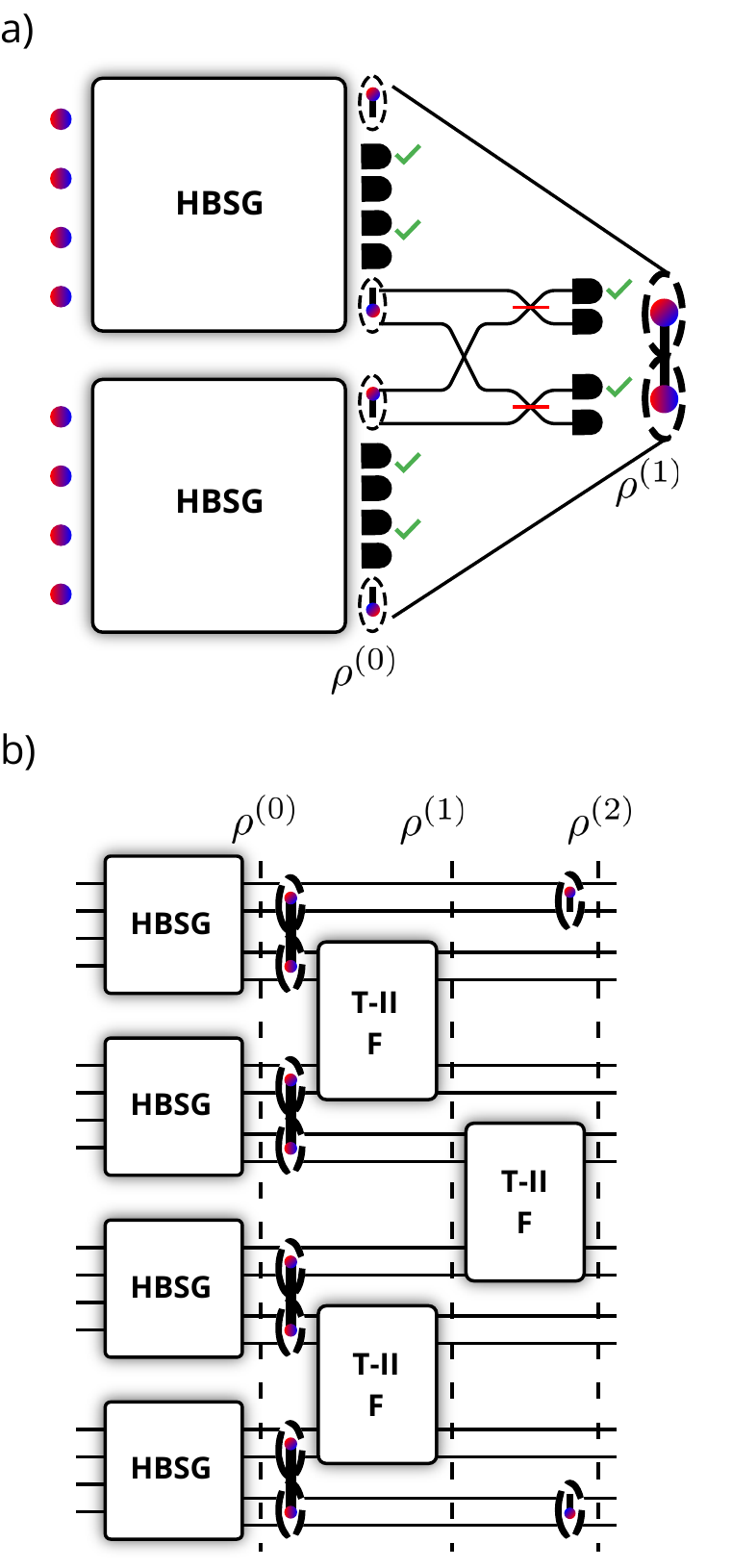}
	\caption{\label{fig:fusioncircuits} \textbf{a)} 8P16M circuit consisting of two 4P8M HBSGs and an appended Type-II fusion operation. Successful heralding with indistinguishable input photons results in projection onto the $\ket{\Phi^{+}}$ state. \textbf{b)} General concantenated Type-II fusion circuit with HBSGs acting as entangled state resource generators for successive (T-II F) gate operations.}
\end{figure}

Computing the corresponding fidelity $\mathcal{F}_{\tilde{\rho}}$ with ${\ket{\Phi^{+}}}$, we benchmark the logical state degradation for each HBSG with both the RS and OBB models. These results are shown in Fig. \ref{fig:1stQRSOBB} along with comparative data from Fig. \ref{fig:2ndQRSOBB}a (washed-out). For non-unity visibilities we observe that $\mathcal{F}_{\tilde{\rho}} < \mathcal{F}_{\rho}\; \forall \; \mathcal{V}$. Occupation of the non-computational space in both the symmetric and anti-symmetric subspaces is accounted for faithfully for $\tilde{\rho}$ states in the Schur-Weyl basis. For $\rho$ (where postselection has occurred) the computational anti-symmetric contributions are incorrectly assigned to the computational symmetric subspace and manifest as decoherence. The probability of being in the computational subspace for the full Fock reconstruction can be equated to the equivalent terms in $\tilde{\rho}$, defined across the symmetric and anti-symmetric two-photon subspaces. The disparity between $\rho$ and $\tilde{\rho}$ persists even in the high visibility regime (see insets Fig. \ref{fig:1stQRSOBB}), suggesting that degradation due to bosonic leakage is likely to impact realistic LOQC implementations. Whilst the 5P5M HBSG is still the most robust to imperfections, relative circuit performance for tomographic reconstruction in this regime is not strictly maintained, with the 4P6M circuit the most sensitive in the first quantized paradigm. These results encapsulate the underrepresentation of state degradation due to partial distinguishability and non-computational leakage when performing postselected state reconstruction.

\section{Optical Type-II Fusion}\label{opint}

The small entangled states modelled can be used as seed states to produce larger qubit states which are vital resources in LOQC architectures \cite{fusion-based-qc} or repeater networks \cite{azuma2015all}. Growing these larger states can be achieved via projective entangling measurements, known as fusion gates. There exist two fusion gate variants which differ on the number of photons measured, either a single photon (Type-I) or both (Type-II). By measuring both input photons, the Type-II gate has an increased tolerance to loss over Type-I, and is therefore the focus of the following section. Type-II fusion can be implemented by a partial BSM (as introduced in Sec. \ref{opticalquantumtelep}).

For our purposes of comparison, we perform fusion on two Bell states generated by HBSGs. On successful state heralding, the remaining qubits are left in a Bell state. Whilst these initial resource states do not permit growth of larger cluster states, modelling the gate in this manner allows us to isolate the gate operation. We combine two 4P8M HBSG circuits in parallel, with the projective Type-II fusion measurement appended to one output qubit from each, resulting (in the error-free case) in projection onto $\ket{\Phi^{+}}$ on the unmeasured qubit modes. Whilst the 5P5M HBSG circuit exhibited the highest robustness in Sec. \ref{firstq}, the 4P8M circuit performed comparably and has a lower simulation overhead in terms of photon number. The full 8P16M linear optical circuit is presented in Fig. \ref{fig:fusioncircuits}a, with ideal success probability of $1.2\times10^{-4}$ \cite{successprobkey, losimgit}.


\begin{figure*}
	\centering
	\includegraphics[width=0.95\textwidth]{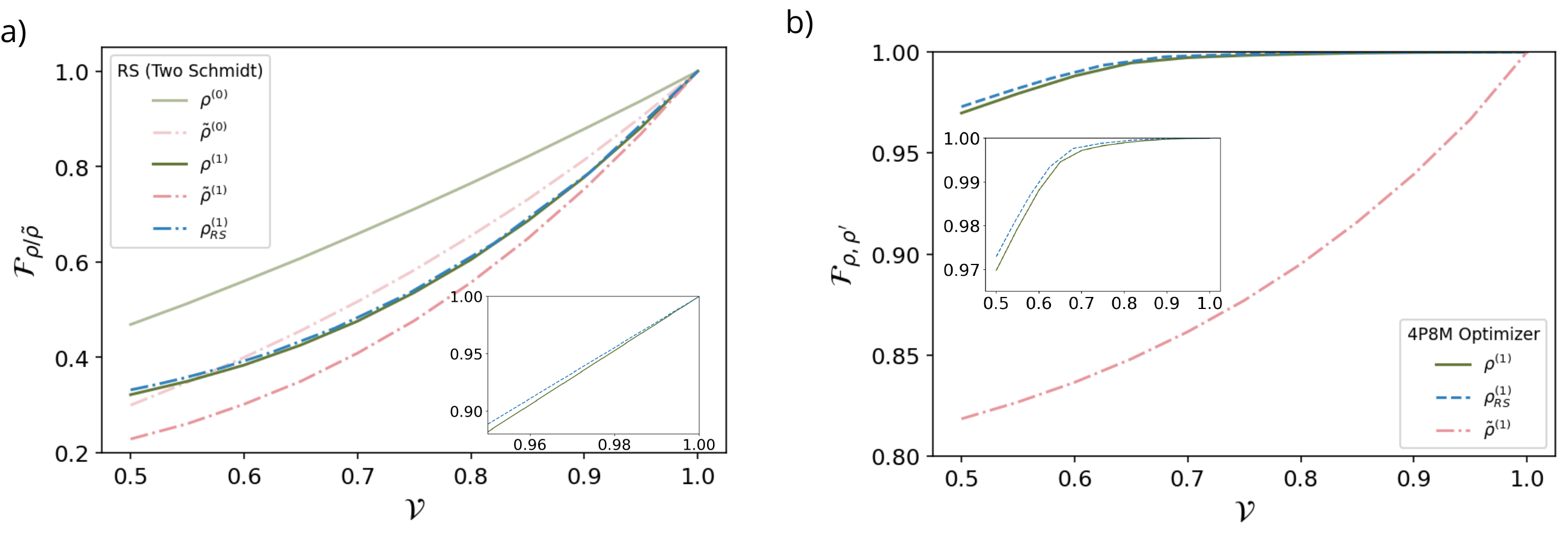}
	\caption{\label{fig:fusionresults}\textbf{a)} Relative state degradation for $\rho$ and $\tilde{\rho}$ seed states and subsequent Type-II gate operation with RS (two Schmidt) photon sources. Inset focuses on post-fusion states for both the analytic expression $\rho^{(1)}_{RS}$ and $\rho^{(1)}$ which permits non-computational leakage at the HBSG step, in the high visibility regime. \textbf{b)} Fidelity of closest state $\rho^{\prime}$ physically accessible from the 4P8M HBSG (optimized) to a post-initial Type-II fusion state $\equiv{\rho^{(1)}_{RS},\rho^{(1)},{\tilde{\rho}}^{(1)}}$ (see Fig. \ref{fig:fusioncircuits}a). Inset again focuses on $\rho^{(1)}_{RS},\rho^{(1)}$ states.}
\end{figure*}

\subsection{Postselected gate operation}
Describing logical state degradation from concatenated fusion operations (see Fig. \ref{fig:fusioncircuits}b.) would ideally take a modular approach, due to overheads in simulating larger linear optical circuits with partially distinguishable photons. For a modular characterization, we attempt to generate a post-HBSG state with the same density matrix as the post-fusion state, i.e. mapping the fusion operation onto the HBSG output state. More specifically, if by tuning the visibility of photons into the HBSG we can recover the post-fusion density matrix, then we use the single 8P16M circuit characterization to simulate the effect of subsequent fusion operations.

If we assume that postselected state tomography is performed at the output of the HBSG then the form of the density matrix (in the computational subspace) can be derived analytically \cite{sparrow_quantum_2017}. For the OBB distinguishability model, we see that the post-fusion density matrix is equivalent to the post-HBSG density matrix with a visibility remapping of $\mathcal{V}\mapsto\mathcal{V}^2$. For the specific case of fusing two Bell states, the fusion gate has a simple operation on the density matrix and concatenated fusion gates with OBB photons can be modelled through the visibility relabelling procedure. For the RS distinguishability model, however, we observe that this simple relationship breaks down and, in fact, it is impossible to find a visibility change that leaves the density matrices the same \cite{fixedschmidtdim}. This suggests that there is some action from the fusion gate that is not fully captured by the HSBG, implying that we cannot simply model multiple gates in this way. The details of the analytic state derivations for postselected Type-II fusion with RS and OBB input photons are provided in Appendix \ref{rsobbanalyticmatrices}. An important caveat here is that these state representations have not considered leakage into the wider non-computational space, thus irrespective of whether the modular characterization procedure can be implemented, we need to investigate whether this behaviour extends to seed states characterized in the Schur-Weyl basis.

\subsection{Contamination from non-computational leakage}\label{bosonicleakagecont}
 
To consider contamination from non-computational leakage, we simulate the full 8P16M circuit in the presence of RS sources (two Schmidt) within the CV backend. The circuit characterization quantifying state degradation is provided in Fig. \ref{fig:fusionresults}a. Here both the second quantized $\rho$ and first quantized $\tilde{\rho}$ states are characterized, at the post-HBSG and post-fusion stages denoted by the superscript labels $(0),(1)$ respectively (see Fig. \ref{fig:fusioncircuits}a). For completion, we also plot the postselected analytic $\rho_{RS}$ state fidelity. For $\rho, \tilde{\rho}$ state reconstruction we required the use of the \textit{BlueCrystal 4} High Performance Computing cluster, as these simulations are not tractable on a standard laptop. 
 Relative state degradation between $\rho^{(0)}$ and $\rho^{(1)}$ is larger than that between the corresponding $\tilde{\rho}$ states. In the Fig. \ref{fig:fusionresults}a inset we show both the post-fusion $\rho^{(1)}$ state fidelity and the analytically calculated $\rho^{(1)}_{RS}$, in the high visibility regime. The discrepancy here is due to the propagation of non-computational terms into the fusion gate which contaminate the output state, but are not accounted for in the derivation of $\rho^{}_{RS}$. This leakage into the wider bosonic space at the intermediate step is accounted for in the CV backend when simulating the full 8P16M circuit. We discuss these error mechanisms for first quantized $\tilde{\rho}$ states, quantifying leakage contributions from both the symmetric and anti-symmetric two-photon subspaces in Appendix \ref{leakage-app}. An important consideration is that in this example, the fusion gate only acts on one qubit from each $\rho^{(0)}$ seed resource. More general resource generation protocols may require a higher depth of gate operations, resulting in additional unaccounted logical state contamination \cite{adcock_hard_2018}. 
 We also include the first quantized Schur state characterization (denoted as $\tilde{\rho}^{(1)}$) which provides insight into the non-computational terms remaining post-fusion.


Previously, we have seen that the analytical form of the post-HBSG state does not allow a one-to-one visibility remapping between post-HBSG and post-fusion states in the postselected regime with RS photons. To consolidate this belief, we attempt to generate the post-fusion Bell states by tuning the HBSG input photon visibility. To find the closest state $\rho^{\prime}$ to the target post-fusion Bell state (referred to now as $\rho$), we maximize the fidelity $\mathcal{F}_{\rho,\rho^{\prime}}$. For the analytic density matrices, we employ a Nelder-Mead optimization algorithm. For the first and second quantized density matrices, we leverage the CV backend and perform a brute force parameter sweep. We plot the maximum fidelity for each of these settings as a function of the input photon visibility to the 8P16M circuit in Fig. \ref{fig:fusionresults}b. As expected, we see for all non-unity visibilities we cannot ever recover the exact target density matrix.
This suggests that part of the photonic state generated after the fusion operation is not reproducable from HBSG \cite{fixedschmidtdim}. Postselection results in the loss of critical diagnostic error information for photonic states in the presence of partially distinguishable photons. A faithful characterization of the noise process in this setting for photonic states after successive fusions is not permitted with a modular characterization approach. It is worth noting here, we do not consider explicit fusion failures, which result in incorrect heralding patterns. These are heralded errors and can be handled by multiplexing or subsequent quantum error correction protocols \cite{fusion-based-qc}.

\section*{Discussion}
In this work we have analyzed heralded linear optical circuits in the presence of physical errors. Leveraging the flexibility of a CV simulation framework, we have shown that the 5P5M HBSG scheme has the highest robustness to partial distinguishability errors for near-ideal photons, when constraining the state to the computational subspace. Efficient optical quantum teleportation depends strongly on the interference scheme used to generate the entangled resource consumed in the protocol, exhibiting similar relative robustness to state generation schemes. Addressing the limitations of postselected state reconstruction in capturing the leakage of heralded resources into the wider non-computational bosonic space, we have utilised the first quantized particle description of photons to model these effects more faithfully. Again, the 5P5M scheme exhibits the highest robustness to these errors for near-ideal photons. The first quantized state representation can be used as a diagnostic tool to elucidate photonic states in computational primitives. Characterization of the Type-II fusion gate in the presence of partially distinguishable photons suggests that true state degradation is underrepresented when we perform postselection. Here, we take a first step in revealing the true noise models associated with heralded partially distinguishable resource states in linear optical circuits.

These insights may lead to modifications in proposed architectures for LOQC to mitigate for such errors, be that from the mechanisms used for resource state generation or subsequent gate operations. Moving forward, tools elucidating the contaminative effects of imperfect photon interference should be developed further. Due to computational overheads, our Type-II fusion simulations used two Bell states as resources to probe non-computational leakage. Typical resource states in FBQC such as the 4-star and 6-ring graph states require more complex generation circuits with Bell and 3-GHZ seed states as initial resources. Whilst the noise in these processes will be bounded, and errors are intrinsically local due to the FBQC construction \cite{fusion-based-qc}, we would need to optimize the CV backend to handle increasing photon number, even for modular characterization. One natural extension is to consider the practical robustness of photon distillation schemes to such noise models \cite{sparrow_quantum_2017, marshall_distillation_2022}. It may also be fruitful to explore if unitary averaging protocols which have theoretically demonstrated improvements in the fidelity of cluster states from Type-II fusion can provide increased resilience to such errors \cite{singh_optical_2022}. As partial distinguishability and resulting non-computational leakage are faithfully modelled in the first quantized state representation, the understanding of photonic entangling gates in this setting should continue to be developed. This could potentially pave the way to reliable modular resource state characterization tools. More broadly, a vital next step is in mapping the discussed leakage mechanisms to code-level errors for explicit relation to fault-tolerant photonic quantum computing architectures. 




\begin{footnotesize}
\noindent
\\
\noindent\textbf{Acknowledgements}
\noindent The authors would like to thank T. J. Bell, J. F. F. Bulmer, A. Ma\"\i{nos} and P. S. Turner for useful discussions.\\ The authors acknowledge support from the Engineering and Physical Sciences Research Council (EPSRC) Hub in Quantum Computing and Simulation (EP/T001062/1). Fellowship support from EPSRC is acknowledged by A.L (EP/N003470/1). R. D. S is supported by the Quantum Engineering Centre for Doctoral Training, EPSRC grant EP/LO15730/1. Part of this work was carried out using the computational facilities of the Advanced Computing Research Centre, University of Bristol - \url{http://www.bristol.ac.uk/acrc/}.
\end{footnotesize}
\bibliography{biblo.bib}

\begin{thebibliography}{64}
\bibitem [{suc()}]{successprobkey}%
  \BibitemOpen
  \href@noop {} {}\bibinfo {note} {The ideal circuit success probabilities we
  quote in the main text are computed by considering a fixed state heralding
  pattern which generates our target resource state ((up to output mode
  permutations) see Fig. \ref{fig:bsgs}). For the 4P8M HBSG there are multiple
  patterns which generate maximally entangled Bell states which contribute to
  the overall success probability of the HBSG circuits. These calculations
  utilise the LOSim Python module (see Ref. \cite{losimgit}).}\BibitemShut
  {Stop}%
\bibitem [{fix()}]{fixedschmidtdim}%
  \BibitemOpen
  \href@noop {} {}\bibinfo {note} {Here we have fixed the internal state
  dimension in attempts to recover the post-fusion state in simulation to
  $d=2$, i.e. two Schmidt terms. The region of interest for state generation is
  at high photon visibility $\mathcal{V}\sim{1}$, where we only expect two
  non-zero Schmidt terms with one dominant.}\BibitemShut {Stop}%
\end{thebibliography}

\newpage
\clearpage
\onecolumngrid
\appendix

\section{3-GHZ state generation}\label{ghzapp}
\begin{figure}[!htbp]
	\centering
	\includegraphics[width=0.33\textwidth]{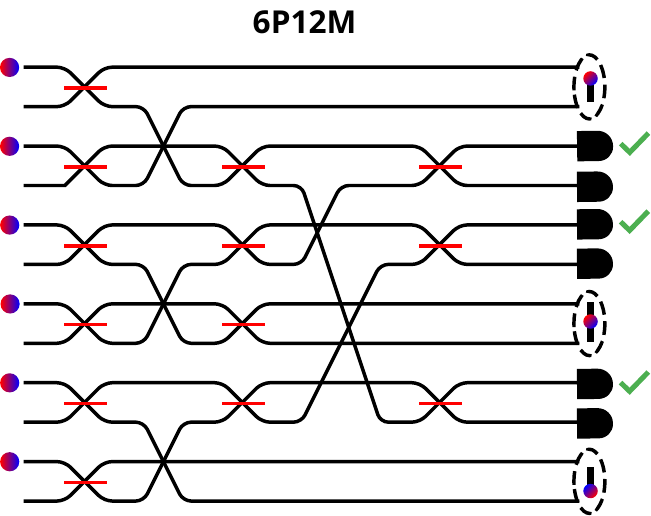}
	\caption{\label{fig:ghzint}6P12M heralded scheme for GHZ state generation.}
\end{figure}

$3$-GHZ states can be efficiently generated with six single photons. Generalising the 4P8M HBSG scheme to six photons across twelve modes (6P12M) and successfully heralding on three photons (with success probability $3.9\times10^{-3}$) \cite{successprobkey, losimgit}, we obtain a state locally equivalent to the $3$-GHZ state (see Fig. \ref{fig:ghzint}):
\begin{equation}\label{ghzstate}
\ket{3-GHZ}=\frac{1}{\sqrt{2}}(\ket{000}_L+\ket{111}_L).
\end{equation}
Unlike with Bell states as fusion resources, subsequent fusions with $3$-GHZ states permit the creation of 4-GHZ (and larger) states, allowing for the growth of photonic cluster states. A cluster state structure of \textit{trees} that can tolerate up to $50\%$ qubit loss was introduced in \cite{Varnava_2006}. The high threshold is a result of the error model being restricted to qubit loss as the dominant mechanism. Further work \cite{varnava_how_2008} showed that efficient growth was also possible for non-unity efficiency photon sources and detectors. More specifically, it is stated that efficient LOQC is possible when $\eta_s\eta_h>2/3$, where $\eta_s,\eta_h$ are the source and detector efficiencies respectively. The loss tolerant protocol of \cite{Varnava_2006} is robust to uncorrelated loss where the cluster states generated are of an ID (independently-degraded) form. The ID-GHZ state structure leads to the $\eta_s\eta_h>2/3$ result stated above. Here we simulate the 6P12M circuit for generating $3$-GHZ states, reconstructing the state with the CV backend to investigate the logical state degradation for RS and OBB photon sources. We also model this circuit with indistinguishable photons, under the constraint of photon loss and threshold detection as discussed for HBSGs in Appendix \ref{threshandloss}. Here we see if the ID-GHZ state form is retained for more general source model constraints with indistinguishable photons.

\FloatBarrier

\begin{figure}[!htbp]
	\centering
	\includegraphics[width=0.9\textwidth]{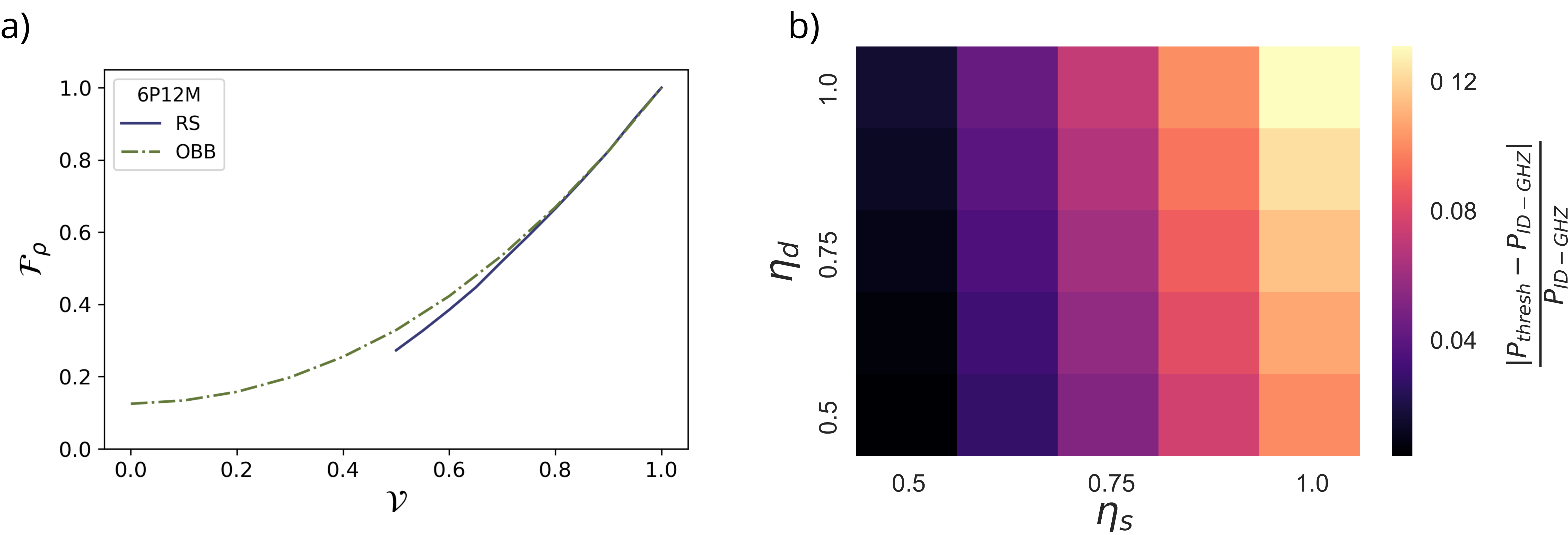}
	\caption{\label{fig:ghzplots}Performance benchmarking of the 6P12M circuit. \textbf{a)} Logical 3-GHZ state fidelities $\mathcal{F}_{\rho}$ as a function of the pairwise photon visibility, $\mathcal{V}$ for RS and OBB sources. \textbf{b)} Heatmap of $|P_{thr}-P_{ID-GHZ}|/P_{ID-GHZ}$ with indistinguishable photons in the presence of threshold detectors and an extra unwanted photon.}
\end{figure}

In Fig. \ref{fig:ghzplots}a we report $\mathcal{F}_{\rho}$ for both RS and OBB sources, displaying equivalence between RS and OBB in the high visibility regime, similar to the 4P8M HBSG circuit, of which this circuit is a natural extension of (in terms of interferometry). In \cite{varnava_how_2008}, the photonic states injected into the 6P12M interferometer have the form $\rho=\eta_s\ket{1}\bra{1}+(1-\eta_s)\ket{0}\bra{0}$, which is an appropriate assumption for lossy, otherwise ideal deterministic single photon emitters. Imperfect QD single photon emitters or heralded spontaneous photon sources will lead to the possibility of multi-photon contamination. Attempting this state generation protocol, perhaps using multiplexing with threshold detection, can lead to such terms degrading the quality of the output state. The general expression for the ID-GHZ state that is generated (conditional on state heralding of three photons) in the 6P12M circuit is given in the Supplementary Info. of \cite{varnava_how_2008}, which contains $0(\ket{vac})-3(\ket{GHZ})$-photon components. We focus on the amplitude of the three-photon $\ket{GHZ}\bra{GHZ}_{ID-GHZ}$ component which is given by
\begin{equation}
 (1-f)^3\ket{GHZ}\bra{GHZ}_{ID-GHZ}, 
\end{equation}
parameterised in terms of the \textit{loss rate}, $f$, as
\begin{equation}
f=1-{\frac{\eta_s}{2-\eta_s\eta_h}}.
\end{equation}

We compute the population of the $\ket{GHZ}\bra{GHZ}_{thr}$ component relative to lower photon-number components using the CV backend as $P_{thr}$, now permitting contamination from an additional photon as described in Sec. \ref{threshandloss} for HBSGs. $|P_{thr}-P_{ID-GHZ}|/P_{ID-GHZ}$ defines the difference between this population and the ideal population of the $\ket{GHZ}\bra{GHZ}_{ID-GHZ}$ term when injecting $\rho=\eta_s\ket{1}\bra{1}+(1-\eta_s)\ket{0}\bra{0}$ photonic states into the 6P12M interferometer. This data is given in Fig. \ref{fig:ghzplots}b for varying $\eta_s,\eta_h$ values, with the largest disparity given for unity-efficiency state and source herald detectors.

\FloatBarrier

\section{Continuous-variable (CV) preliminaries}\label{cv-app}
In the CV formalism we can define a quantum state, $\hat{\varrho}$, over the set of modes, $l$, through the associated Wigner function
\begin{equation}
W(\Vec{\alpha};\hat{\varrho})=\int\frac{d\Vec{\xi}}{\pi^{2l}}\Tr[\,\hat{\varrho}\hat{D}(\Vec{\xi})]\,\exp({\Vec{\alpha}}^{T}\Omega\Vec{\xi}),
\end{equation}
where $\Vec{\alpha},\Vec{\xi}$ are complex bivectors, $\hat{D}(\xi)=\exp(\Vec{\xi}^T\Omega\hat{\zeta})$ is the displacement operator ($\Omega$, the symplectic form) and $\hat{\zeta}_j$ is the creation-annihilation bivector for $j=1,\ldots,l$. $\hat{\zeta}_j=\hat{a}_j$ and $\hat{\zeta}_{l+j}={\hat{a}}_j^{\dag}$ satisfy the canonical commutation relations $[\,\hat{a}_i,\hat{a}_j]\,=0$ and $[\,\hat{a}_i,\hat{a}_j^\dag]\,=\delta_{ij}$.

We can characterize a multimode Gaussian state $\hat{\varrho}$ through its first and second moments; the displacement vector $\Vec{\beta}$ with
\begin{equation}
{\Vec{\beta}}_j=\Tr[\,\hat{\varrho}\hat{\zeta}_j]\,
\end{equation}
and covariance matrix, $\sigma$ with elements
\begin{equation}
\sigma_{jk}={\Tr[\,\hat{\varrho}\{\hat{\zeta}_j,\hat{\zeta}_k^\dag\}]\,}/2-{\Vec{\beta}}_j{\Vec{\beta}}_k^*.
\end{equation}

Shown in \cite{quesada_simulating_2019}, the following quantities:
\begin{equation}
X=\mqty[0 & \mathbbm{1}_l \\ \mathbbm{1}_l & 0],
\end{equation}
\begin{equation}
\sigma_q=\sigma+\frac{1}{2}\mathbbm{1}_{2l},
\end{equation}
\begin{equation}
T=\frac{\exp(-\frac{1}{2}{\Vec{\beta}}^\dag\sigma_{Q}^{-1}\Vec{\beta})}{\sqrt{\det(\sigma_q)\prod_{s=1}^{l}n_{s}!m_{s}!}},
\end{equation}
lead to the expression for computing Fock matrix elements of $\hat{\varrho}$ as
\begin{equation}
\matrixel{\textbf{m}}{\hat{\varrho}}{\textbf{n}}=T \times \text{lhaf}(\Tilde{A}).
\end{equation}
Elements of $\Tilde{A}$ can be defined as
\begin{equation}
\tilde{A}_{i, j}= \begin{cases}\bar{A}_{i, j} & \text { if } i \neq j \\ \bar{\gamma}_i & \text { if } i=j\end{cases},
\end{equation}
with 
\begin{equation}
A=X(\mathbbm{1}_{2l}-\sigma_Q^-1),
\end{equation}
\begin{equation}
\gamma^T=\beta^{\dag}\sigma_Q^-1.
\end{equation}
The number of steps to perform the lhaf calculation is $O(lAG^{2l})$ where $G$ is a geometric mean quantity defined in \cite{quesada_simulating_2019}.

\section{CV tomographic reconstruction}\label{cvtomo}
To compute the expectation values of Eq. \ref{expvalstomography} with the CV backend, we can take the spectral decomposition of Hermitian operators, $G_i$, as
\begin{equation}
G_i=\sum_{j=0}^{d-1}\lambda_j^{(i)}\ket{v_j^{(i)}}\bra{v_j^{(i)}},
\end{equation}
where $\left\{\ket{v_j^{(i)}}\right\}$ are the corresponding eigenvectors and $\lambda_j^{(i)}$ are the associated eigenvalues. Rotating from the eigenvector basis to a computational qudit basis $\{\ket{j}\}$ is permitted through an optical unitary, $U_i$, of the form
\begin{equation}
U_i=\sum_{j=0}^{d-1}\ket{j}\bra{v_j}.
\end{equation}
This rotation is carried out by appending $U_i$ to the modes representing each qudit of the computational subspace state, post transfer matrix. The probability, $P(k,l..|i,j..)$, of detecting photons in modes $\{i,j,..\}$ can be given by
\begin{equation}\label{tomoprobs}
\Tr[\,\rho\left(\ket{v_k^{(i)}}\otimes\ket{v_l^{(j)}}\otimes..\right)\left(\bra{v_k^{(i)}}\otimes\bra{v_l^{(j)}}\otimes..\right)]\,,
\end{equation}
where probabilities are computed through expressions of the form of Eq. \ref{matrixelem} for $\textbf{m}=\textbf{n}$. From Eq. \ref{tomoprobs}, we can define expectation values $\langle{G_{i1}}\otimes...\otimes{G_{in}}\rangle$ as
\begin{equation}
\sum_{k,l..}\lambda_{k}^{(i)}\lambda_{l}^{(j)}..\times{P(k,l..|i,j..)}.
\end{equation}

\section{Filtering of source firings for efficient state reconstruction}\label{sourcefiringlogic}
When reconstructing states from linear optical entanglement generation circuits with RS photon sources, there are equivalent valid source herald patterns that must be independently simulated, with one simulation prescribed to a single \textit{source firing}. In the RS model the number of source firings, $N_{SF}$, is related to the number of input photons, $|\chi|$, and Schmidt number $K$ as $N_{SF}=K^{|\chi|}$. A given source firing constrains the distribution of photons across Schmidt modes, allowing us to filter out a subset of state herald patterns. By summing over photon occupations $\textbf{m},\textbf{n}$ with the same number of photons in each Schmidt mode, we reduce the number of lhaf calculations required for a specific Fock matrix element of the non-Gaussian output state (see Eq. \ref{matrixelem}). This allows for more time-efficient full Fock space state reconstruction when $K\geq2$ compared to not filtering out any source firings.

\section{Threshold detectors and photon loss}\label{threshandloss}
For the proposed HBSG circuits, we have so far considered PNRDs that measure the number of photons in an optical mode. In addition to photon distinguishability, we can assess the impact of threshold detection and photon loss on generated Bell state fidelities. Photon loss can occur due to photon scattering, absorption, or inefficiences in detection. Losses can be modelled through a beam splitter with transmission coefficient $\eta$, where transmitted photons are detected with an ideal photon detector and reflected photons are considered lost. The modes supporting photonic qubits may also suffer from loss, leading to false positives which are not resolved by detection apparatus. Indistinguishable photons with source (state) detector efficiencies $\eta_{s(h)}$, are modelled by applying Gaussian single-mode loss channels $\hat{a}\mapsto\sqrt{\eta_{s(h)}}\hat{a}+\sqrt{1-\eta_{s(h)}}\hat{b}$ to the source (state) herald modes pre(post)-transfer matrix, with $\hat{b}$ representing an ancillary vacuum mode. When loss is present, multi-photon contamination from parametric photon sources is important, as higher-order photon terms are not perfectly heralded as they are in the lossless PNRD case.

\begin{figure}[!htbp]
	\centering
	\includegraphics[width=0.45\textwidth]{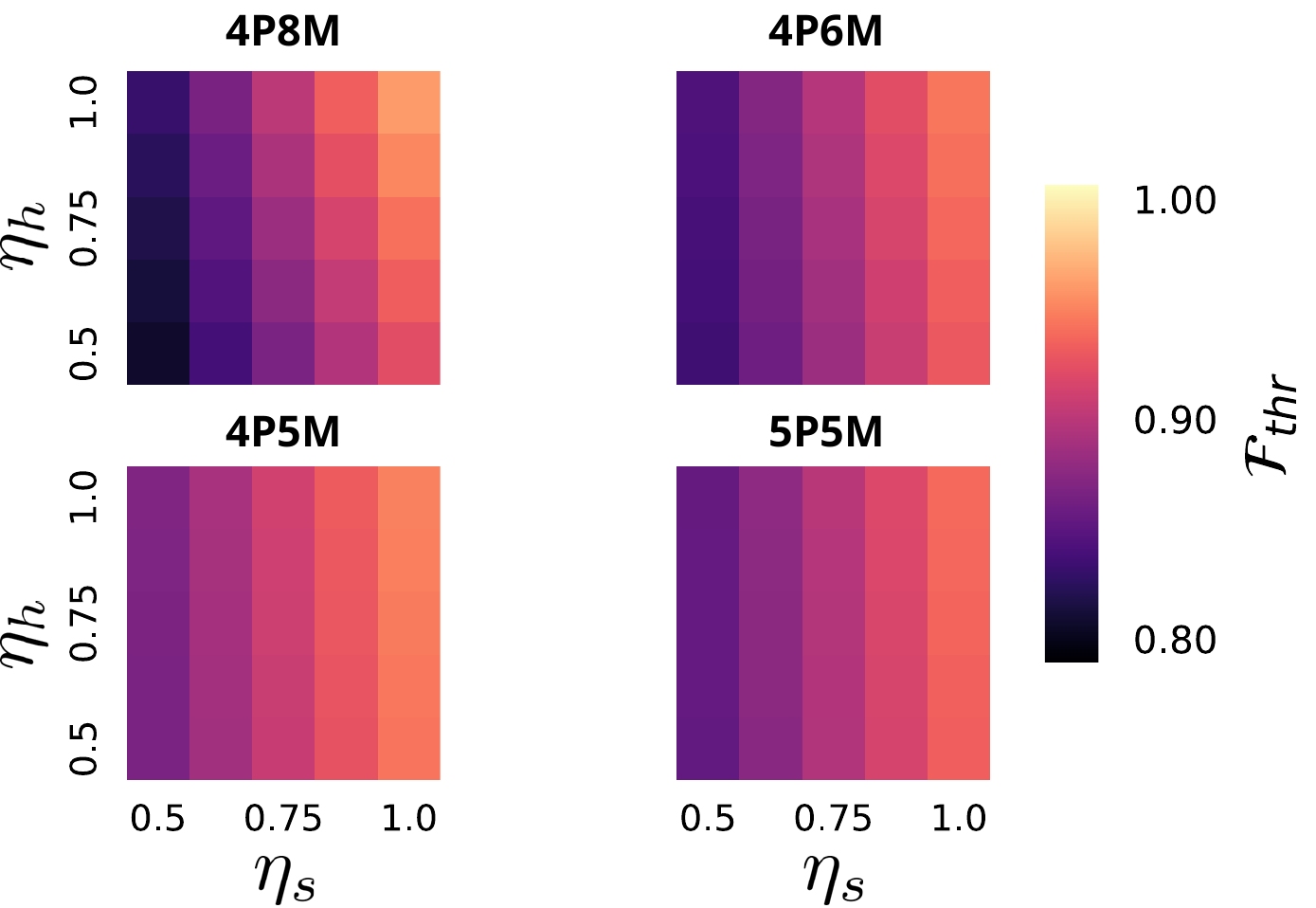}
	\caption{\label{fig:threshbsgs}Heatmap of $\mathcal{F}_{thr}$ for each HBSG circuit with indistinguishable photons in the presence of threshold detectors and an extra unwanted photon.}
\end{figure}
\FloatBarrier

We model the performance of each HBSG by computing the fidelity, $\mathcal{F}_{thr}$ for varying $\eta_s,\eta_h$ values in Fig. \ref{fig:threshbsgs}. To express the ignorance of threshold detection to higher-order contributions, we adapt the CV backend logic to consider terms with higher photon occupations and incoherently sum over these and valid logical photon mode occupations. We truncate by only including cases where one of the sources produces two pairs of photons (squeezing value of $\kappa=0.2$). This choice is well-motivated as realistic photon sources will operate in the low-squeezing limit which means higher-order photon terms are exponentially suppressed. Our simulation always considers the emission of an additional photon meaning that we never expect to see unity $\mathcal{F}_{thr}$ values. The maximum $\mathcal{F}_{thr}$ value is $0.96$ for the 4P8M circuit with $\eta_s=\eta_h=1$. The 4P6M and 4P8M circuits are affected the most by reduction in $\eta_s$ for a fixed $\eta_h$ value. Logical state fidelities are more sensitive to inefficiencies in the source herald detectors before the transfer matrix than on the state herald detectors. Whilst state heralding comprises a subset of the output circuit modes, each idler photon requires the valid heralding from a signal photon so loss in these modes will then be compounded. For more compact HBSG schemes requiring multi-photon state heralding on a single PNRD in the ideal case (4P5M, 5P5M), the entanglement generation protocol is minimally impacted by inefficiencies in the state herald detectors for a fixed $\eta_s$ value. In Appendix \ref{ghzapp} we analyze the 6P12M GHZ circuit under the same constraints, along with both RS and OBB photon sources with PNRDs as discussed for HBSGs in Sec. \ref{resourcegen}.

\section{HBSG with RS photons for realistic sources}\label{34SchmidtApp}
\begin{figure}[!htbp]
	\centering
	\includegraphics[width=0.9\textwidth]{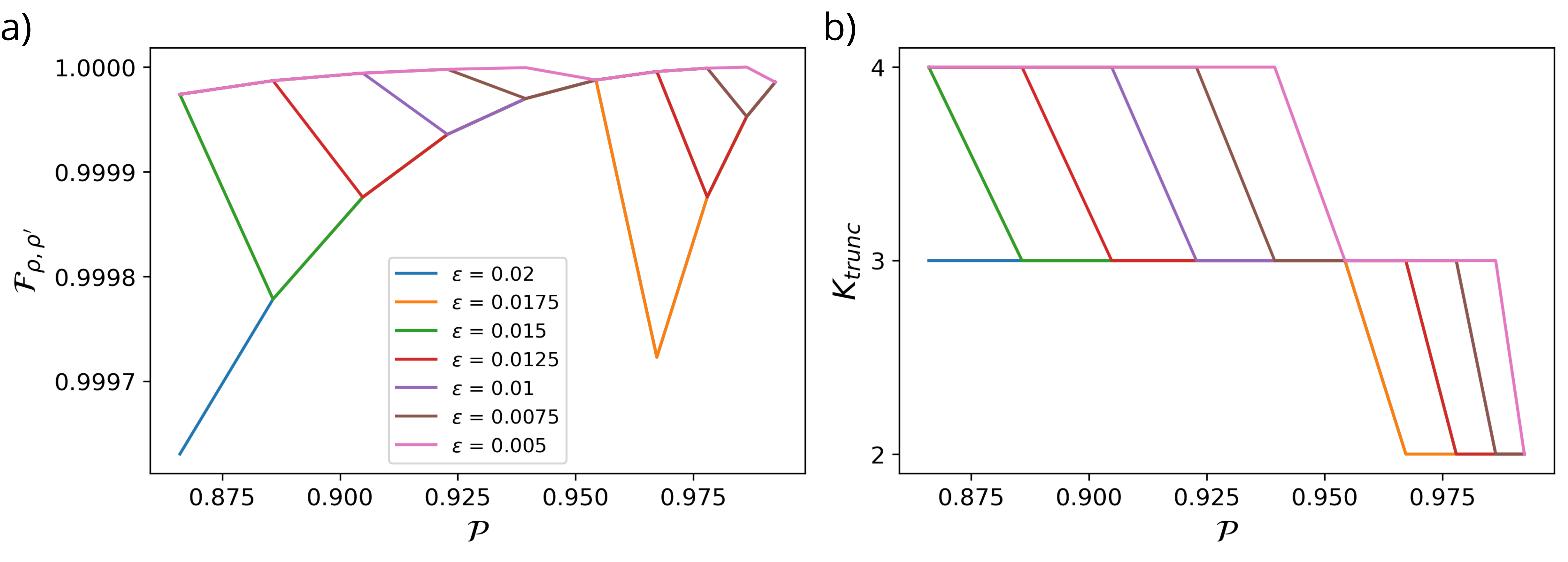}
	\caption{\label{fig:schmidtruncs}\textbf{a)} Fidelity of the SVD state $\mathcal{F}_{\rho,\rho'}$ truncated for different Schmidt coefficient thresholds $\epsilon$ relative to $\rho$, over the purity range $\mathcal{P}$ defined by the filter bandwidth. \textbf{b)} Corresponding number of non-zero Schmidt terms for the state $\rho'$.}
\end{figure}

Here we benchmark the proposed HBSG circuits for a realistic photon source to demonstrate flexibility in the CV backend. Spontaneous processes for generating photons in waveguides are energy constrained, for example in SFWM where the pump bandwidth in the frequency domain, $\omega_{pump}$, satisfies $2\hbar{\omega_{pump}}=\hbar(\omega_{s}+\omega_{i})$. We can optically filter signal and idler bandwidths to constrain photon generation correlations, improving the purity of spontaneous sources. The spectral properties of generated photon pairs are captured by the JSA which quantifies correlation between  signal and idler photons. High purity heralded single photons are related to separability in the JSA. As the JSA can be expressed as a sum of orthogonal modes (see Sec. \ref{distandsources}) then the SVD of the JSA provides us with the set of Schmidt coefficients $\{{s_k}\}$ that are mapped to RS photons of purity $\mathcal{P}$. In general, this decomposition has an infinite number of terms, so to simulate RS photon sources we truncate the SVD of the JSA to provide us with a finite number of Schmidt modes. To ensure this truncation is faithful we define a fidelity $\mathcal{F}_{\rho,\rho'}=\Tr(\sqrt{\rho\sqrt{\rho'}\rho})^2$ where $\rho\equiv{JSA\cdot{JSA^{\dag}}}$ is the untruncated SVD state and $\rho'$ is truncated for a threshold $\epsilon$. We truncate for $s_k \geq \epsilon$, leaving us with a set of non-zero Schmidt coefficients from which we reconstruct $\rho'$. The waveguide simulation assumes silicon with a central wavelength of $\simeq 1550$ nm and pump bandwidth of $400$ GHz.  We compute $\mathcal{F}_{\rho,\rho'}$ in Fig. \ref{fig:schmidtruncs}a for optical filter widths with purity $\mathcal{P}$, for a set of different thresholds. For each $\epsilon$ we report the corresponding number of non-zero Schmidt terms $K_{trunc}$ in Fig. \ref{fig:schmidtruncs}b. Based on $\epsilon=0.005$ providing  $\max(1-\mathcal{F}_{\rho,\rho'}) > \num{1e-5}$ over the whole filter width range, we map the $\{\rho'(\epsilon=0.005)\}$ Schmidt coefficients to RS sources in our CV backend (requiring at most four non-zero Schmidt modes) as described in Sec. \ref{rsmodellingsec}.

\FloatBarrier

\begin{figure}[!htbp]
	\centering
	\includegraphics[width=0.45\textwidth]{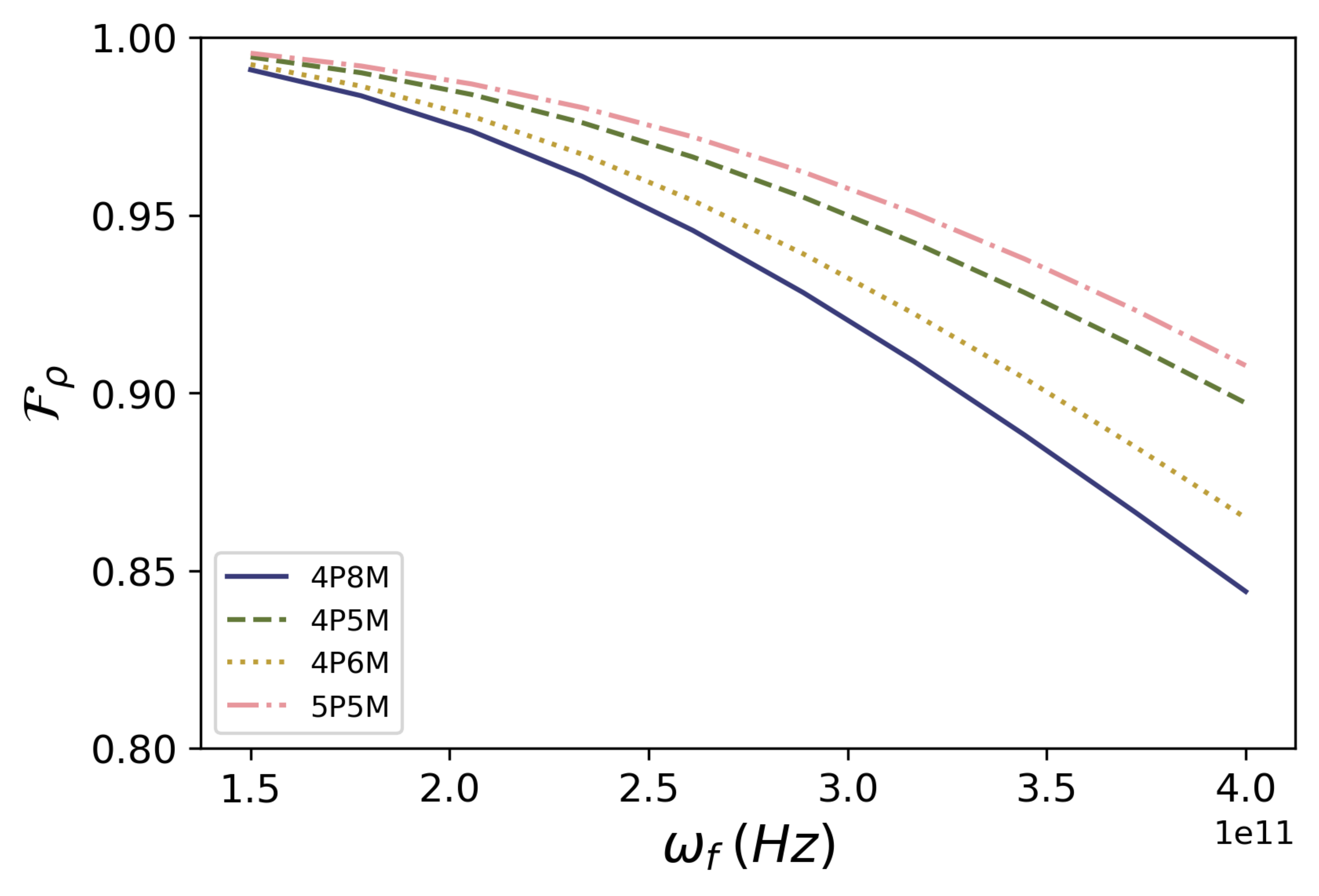}
	\caption{\label{fig:realistichbsgplot} Logical heralded Bell state fidelity (with tomographic state reconstruction) $\mathcal{F}_{\rho}$ for each HBSG circuit, initialized with RS photonic states in $\{\rho'(\epsilon=0.005)\}$.}
\end{figure}

Simulating each HBSG circuit for RS photonic states in $\{\rho'(\epsilon=0.005)\}$ over a filter bandwidth ($\omega_{f}$) range of $150-400$ GHz, we report $\mathcal{F}_{\rho}$ in Fig. \ref{fig:realistichbsgplot}. This filter bandwidth range was chosen to reflect realistic telecommunication wavelength-division multiplexing (WDM) filters. Relative HBSG robustness for RS sources (two Schmidt) simulated in Sec. \ref{resourcegen} is maintained over this filter bandwidth. The maximum discrepancy of $\mathcal{F}_{\rho}$ increases as the filter bandwidth is increased, reducing RS photon purity. An important note here is that for photons with $\mathcal{P} \geq 0.96$, at most only three Schmidt modes are required in simulation to faithfully model these sources to  within $\epsilon=0.005$, with near-ideal spontaneous sources such as those discussed by Paesani et al., \cite{paesani_near-ideal_2020} only requiring two non-zero Schmidt terms.

\FloatBarrier

\section{HBSG state fidelity equivalences (RS and OBB)}\label{rsandobbapp}
Here we display the data from Fig. \ref{fig:2ndQRSOBB}a for each HBSG separately for both proposed distinguishability models, to show that $\mathcal{F}_{\rho}$ is highly constrained by the pairwise photon visibilities $\mathcal{V}$. In \cite{sparrow_quantum_2017} this was shown for the 4P8M HBSG, whereas here we show this property extends to the other HBSG circuits which we analyze.

\begin{figure}[!htbp]
	\centering
	\includegraphics[width=0.9\textwidth]{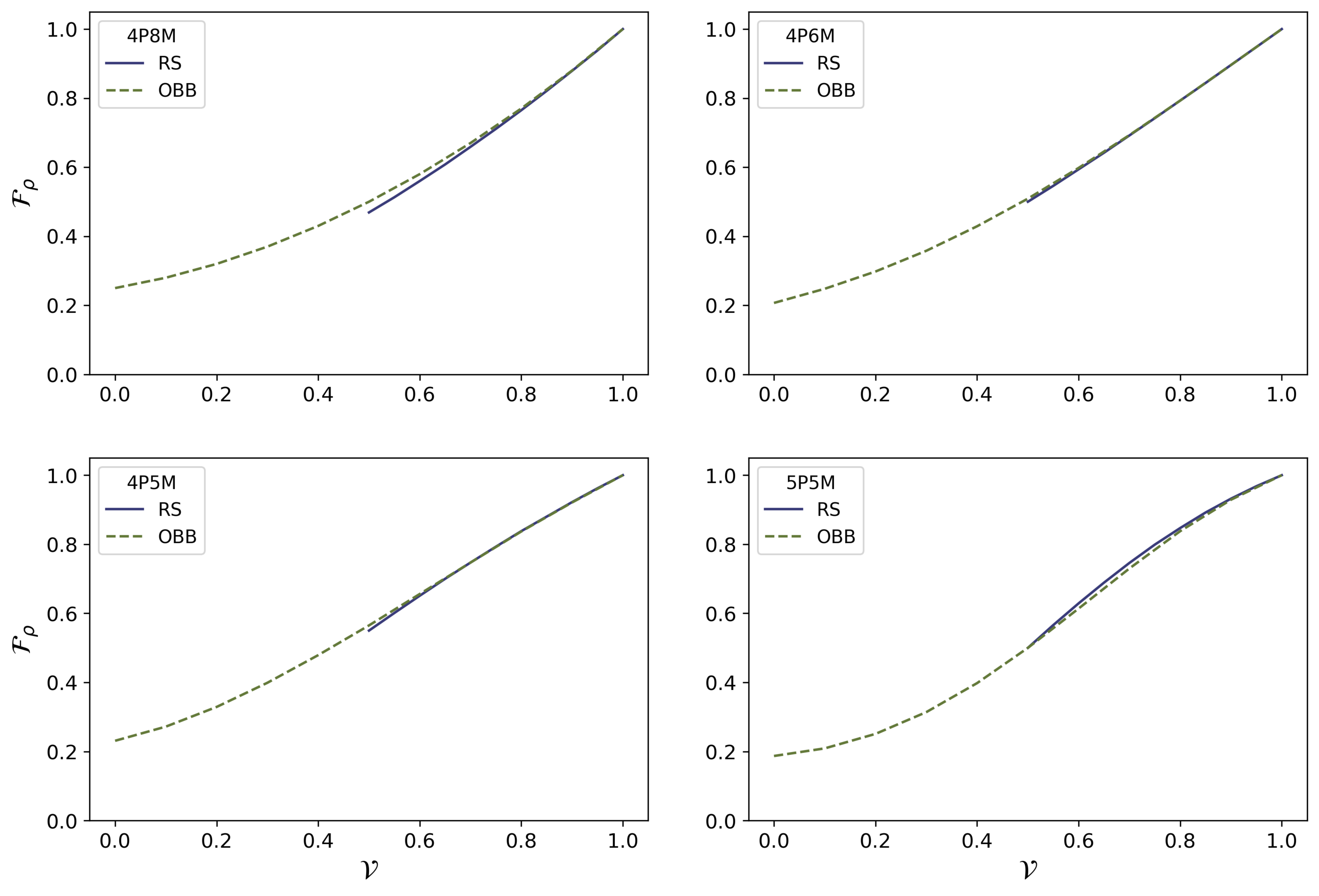}
	\caption{\label{fig:rsvsobbequivs} Relative RS/OBB $\mathcal{F}_{\rho}$ for each HBSG as a function of the pairwise photon visibility $\mathcal{V}$.}
\end{figure}
\FloatBarrier



\section{Optical quantum teleportation of arbitrary single qubit photonic states}\label{supp-telep-info}
Teleportation of the $\ket{+}$ state was analyzed in detail for both RS and OBB sources, with each of the four HBSGs acting as the entangled resource generator, in Sec. \ref{opticalquantumtelep}. Whilst the $\ket{+}$ state comprises the standard initial resource for cluster state generation (pre-CZ gate operation) enabling MBQC \cite{browne-briegel-oneway}, other single qubit states may be of interest in optical quantum teleportation protocols. Here we simulate teleportation protocols (see Fig. \ref{fig:telep}b) for the sets of antipodal single qubit eigenstates of the Pauli matrices, providing coverage of the Bloch sphere. For these simulations the 4P8M HBSG represents the entangled resource generator and photons are of the RS (two Schmidt) form. 

\begin{figure}[!htbp]
	\centering
	\includegraphics[width=0.45\textwidth]{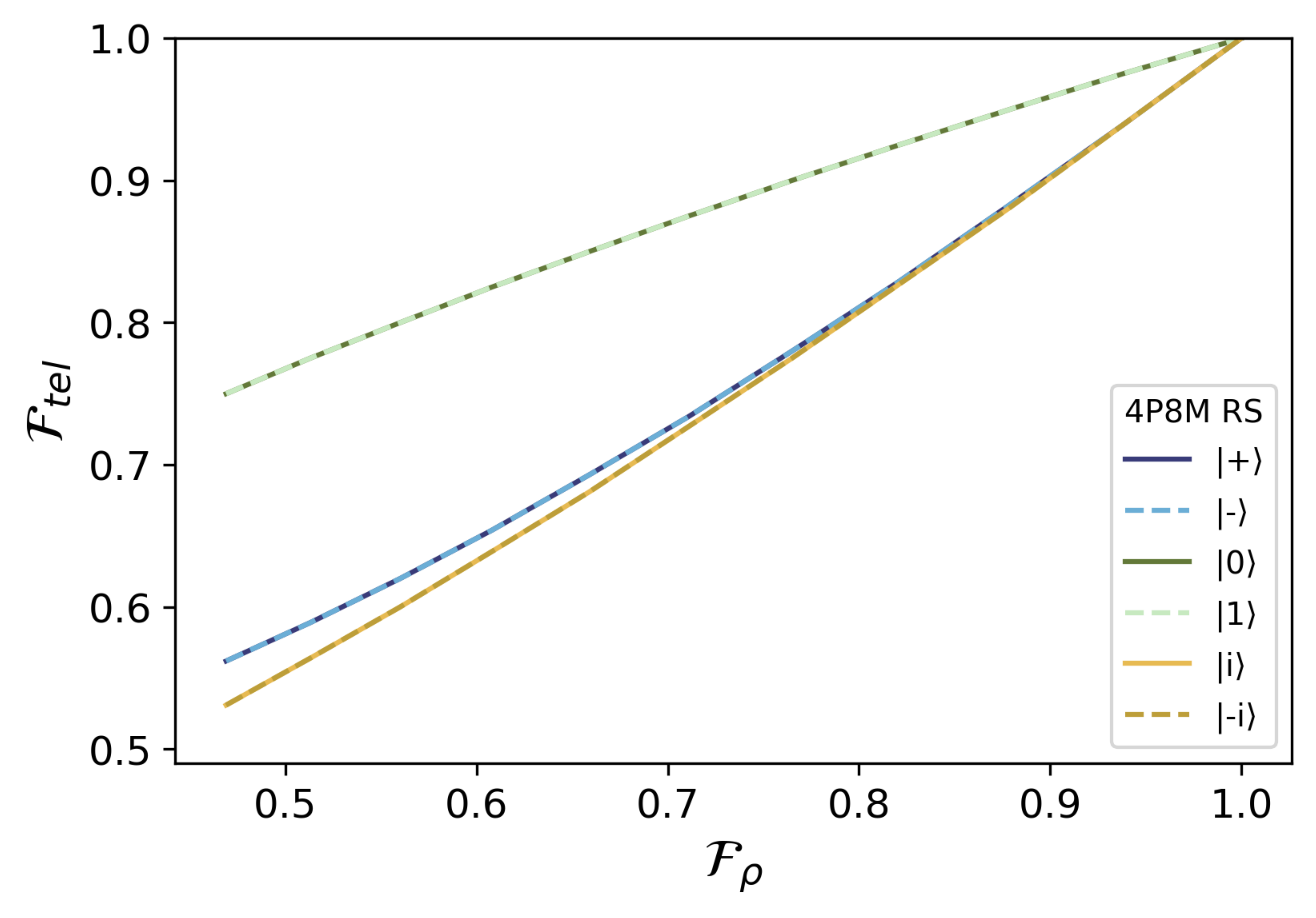}
	\caption{\label{fig:singlequbittelep} Teleported single qubit fidelity $\mathcal{F}_{tel}$ of each of the antipodal single qubit states as a function of $\mathcal{F}_{\rho}$ for RS (two Schmidt) photon sources (4P8M HBSG used as the entangled resource in the protocol).}
\end{figure}
\FloatBarrier

From Fig. \ref{fig:singlequbittelep} we observe that the $\ket{+}$ state $\mathcal{F}_{tel}$ value upper bounds the robustness of the $\sigma_y$ eigenstates $\ket{i}, \ket{-i}$. For each antipodal pair of states, performance is equivalent over the entire state fidelity regime. High $\mathcal{F}_{\rho}$ teleportation equivalence between $\{\ket{+},\ket{-},\ket{i},\ket{-i}\}$ suggests that the discussion concerning implementations of optical quantum teleportation with the $\ket{+}$ state holds more generally, for each of these states.

\section{Particle symmetries and Schur-Weyl duality}\label{SchurWeyl}
For a single particle with two degrees of freedom $(S,L)$ the Hilbert space $\mathcal{H}^{(1)}=\mathcal{H}_S\otimes\mathcal{H}_L$ can be extended for $N$ particles to $\mathcal{H}^{(N)}=(\mathcal{H}^{(1)})^{\otimes{N}}$. Requiring that the wavefunction associated with this $N$-particle space, $\ket{\Psi_D}=\otimes_{j=1}^{N}\ket{\varphi_j}$ obeys bosonic exchange symmetry, we can apply the symmetrisation operator $\hat{\mathcal{S}}$ as
\begin{equation}
\begin{gathered}
\ket{\Psi_S}=\hat{\mathcal{S}}\ket{\Psi_D}\equiv{\vee_{j=1}^{N}}\ket{\varphi_j} \\
=\mathcal{N}_\mathcal{S}\sum\limits_{\sigma\in{S_N}}\ket{\varphi_{\sigma(1)}}\otimes\ket{\varphi_{\sigma(2)}}\otimes...\otimes\ket{\varphi_{\sigma(N)}}.
\end{gathered}
\end{equation}
$\vee$ is the symmetric tensor product, $\mathcal{N}_S$ is a normalisation factor, $S_N$ is the qudit permutation group with elements $\sigma$ and $\ket{\Psi_S}$ is in the symmetrised $N$-particle Hilbert space: $\mathcal{H}_{S}^{(N)}={(\mathcal{H}^{(1)})}^{\vee{N}}$.
Permutation symmetries present in system-label states suggest that we should choose a representation for the two-particle (focused here on Bell states, but $N$-particle more generally) Hilbert space that naturally encapsulates these properties. As we are considering the interference and evolution of photons in linear optical circuits, we also want this representation to behave well under unitary group operations ($SU(2)$ in this case). A strong connection between the permutation group and special unitary group actions is seen through \textit{Schur-Weyl duality} \cite{rowe_dual_2012, bacon_quantum_2005}: for a system of $N$ $d$-dimensional qudits, the associated Hilbert space $\mathcal{H}^{(N)}={(\mathbb{C}^d)}^{\otimes{N}}$ decomposes as
\begin{equation}
    {(\mathbb{C}^d)}^{\otimes{N}}\simeq\bigoplus_{\lambda\vdash{N}}\mathbb{C}^{\{\lambda\}}\otimes\mathbb{C}^{(\lambda)}.
\end{equation}
Here $\mathbb{C}^{\{\lambda\}}$ carries the irrep $\lambda$ of the qudit unitary transformation group, $SU(d)$, $\mathbb{C}^{(\lambda)}$ carries the irrep $\lambda$ of the qudit permutation group, $S_N$, and $\simeq$ indicates a change of basis through the \textit{Schur-Weyl transform}.

The irreps which simultaneously diagonalize the unitary and permutation groups, $\lambda$, can be compactly described through \textit{Young diagrams}. Attributing $N$ qudits with dimension $d$ spanning the particles given degree of freedom, we can construct \textit{Young tableux}, wavefunctions that form a basis for the $N$-particle Hilbert space. For $N=2$ there are two Young diagrams: $\scriptsize\yng(2),\scriptsize\yng(1,1)$. If $d=2$ there are four basis states: the spin triplet
\begin{equation}
\begin{gathered}
    \ket{\scriptsize\young(11)}=\ket{1,1}, \\
    \ket{\scriptsize\young(12)}=\ket{1}\vee\ket{2}=\frac{1}{\sqrt{2}}(\ket{1,2}+\ket{2,1}), \\
    \ket{\scriptsize\young(22)}=\ket{2,2},
\end{gathered}
\end{equation}
and spin singlet
\begin{equation}
    \ket{\scriptsize\young(1,2)}=\frac{1}{\sqrt{2}}(\ket{1,2}-\ket{2,1}).
\end{equation}
These wavefunctions form the symmetric and anti-symmetric bases for the two-particle Hilbert space respectively.

For system-label states, distinguishability in the label allows us to access anti-symmetric irreps, as can be seen for Eq. \ref{distpoleq}:
\begin{equation}
\begin{gathered}
\ket{\psi_{12}^{AB}}=\ket{1,A}\vee\ket{2,B}\\\mapsto\frac{1}{\sqrt{2}}(\ket{\scriptsize\young(12)}\otimes\ket{\scriptsize\young(AB)}+\ket{\scriptsize\young(1,2)}\otimes\ket{\scriptsize\young(A,B)}).
\end{gathered}
\end{equation}
This state, opposed to the case with the same labels on each particle, occupies both $\lambda=(2)$ and $\lambda=(1,1)$.

\section{Schur-Weyl transform implementation in the CV backend}\label{SchurPseudoApp}
We use the QuTip Python library to carry out the Schur-Weyl transform on $\abs{\bar{\chi}}$-particle states reconstructed at the output of our interferometer. Here we provide psuedocode showing the main algorithmic steps for obtaining reduced logical Schur states as part of the CV backend for HBSG circuits.
\begin{algorithm}[H]
  \caption{Schur state reconstruction}
  \label{schuralg}
  \hspace*{\algorithmicindent} \textbf{Input:} Dictionary of full Fock representation $rho-full$ over output state modes, for the set $\bar{\chi}$.\\
  \hspace*{\algorithmicindent} \textbf{Output:} Reduced logical state $\tilde{\rho}$.
   \begin{algorithmic}
   \State $\tilde{\rho} \gets 0$\Comment{Initialise $\tilde{\rho}$}
   \NoDo{}
   \For{$\ket{sp,int}^{\otimes{\abs{\bar{\chi}}}}$ \textbf{in} $rho-full$}
   \State  \ReDo \For{$\bra{sp,int}^{\otimes{\abs{\bar{\chi}}}}$ \textbf{in} $rho-full$}
   \State $\rho_{sym} \gets \hat{S}((\ket{sp,int}\bra{sp,int})^{\otimes{\abs{\bar{\chi}}}})$ \Comment{Symmetrise $\abs{\bar{\chi}}$-photon state}
   \State $\rho_{SW} \gets U_{sp,int}^{\otimes{\abs{\bar{\chi}}}}(\rho_{sym})$ \Comment{Evolve $\rho_{sym}$ through Schur-Weyl transform}
   \State $\rho_{sp} \gets \Tr_{int}({\rho_{SW}})$ \Comment{Trace out internal degree of freedom}
   \State $\tilde{\rho} \mathrel{+}= \rho_{sp}$ \Comment{Update $\tilde{\rho}$}
   \EndFor
   \EndFor
   \end{algorithmic}
\end{algorithm}
\section{4P8M OBB vs RS postselected Type-II fusion}\label{rsobbanalyticmatrices}
In Sec. \ref{opint} we discuss and analyze the Type-II gate operation for both RS (two Schmidt) and OBB photons. The resources for the gate operation are ideally two $\ket{\Phi^{+}}$ Bell states, however in the presence of errors these states are described more generally by $\rho$. For postselected HBSG states, the general structure of $\rho^{(0)}$ was derived in \cite{sparrow_quantum_2017}. This expression depends on the pairwise photon visibility $\mathcal{V}$ and fourth-order trace overlap of the internal photonic state $\varrho$. Here we show the analytic expressions for the post-fusion states $\rho^{(1)}_{OBB/RS}$ resulting from these postselected resources. For OBB photons, simplifying the general $\rho^{(0)}$ expression leads to

\begin{equation}
\rho_{OBB}^{(0)}=\frac{1}{4} \begin{bmatrix}
 1+\mathcal{V} & 0 &  0 &  2\mathcal{V}^2\\
  0 & 1-\mathcal{V} &  0\ &  0\\
  0 & 0 & 1-\mathcal{V} & 0\\
  2\mathcal{V}^2 &  0 &  0 & 1+\mathcal{V}
\end{bmatrix}.
\end{equation}

where the $\ket{01}\bra{10},\ket{10}\bra{01}$ coherences vanish for OBB photons and $\Tr(\varrho^4)=\mathcal{V}^2$. Performing Type-II fusion on two copies of $\rho^{(0)}_{OBB}$ which occupy spatial modes $1,2$ and $3,4$ respectively, preserves terms of $\rho^{(0)}_{OBB}\otimes{\rho^{(0)}_{OBB}}$ as

\begin{equation}
\begin{gathered}
\frac{1}{16}((\alpha\ket{00}\bra{00}+\alpha\ket{11}\bra{11}+\beta\ket{01}\bra{01}+\beta\ket{10}\bra{10}+\gamma\ket{00}\bra{11}+\gamma\ket{11}\bra{00})\\\otimes\\{(\alpha\ket{00}\bra{00}+\alpha\ket{11}\bra{11}+\beta\ket{01}\bra{01}+\beta\ket{10}\bra{10}+\gamma\ket{00}\bra{11}+\gamma\ket{11}\bra{00}))}.
\end{gathered}
\end{equation}

Here we simplify coefficients as $\alpha=1+\mathcal{V}$, $\beta=1-\mathcal{V}$ and $\gamma=2\mathcal{V}^2$. Expanding and grouping terms leads to

\begin{equation}
\begin{gathered}
\frac{1}{16}((\alpha^2+\beta^2)(\ket{00}\bra{00}+\ket{11}\bra{11})+2\alpha\beta(\ket{01}\bra{01}+\ket{10}\bra{10})+\gamma^2(\ket{00}\bra{11}+\ket{11}\bra{00})).
\end{gathered}
\end{equation}

We can then express the output state as

\begin{equation}
\rho_{OBB}^{(1)}=\frac{1}{4}\begin{bmatrix}
 1+\mathcal{V}^2 & 0 &  0 &  2\mathcal{V}^4\\
  0 & 1-\mathcal{V}^2 &  0 &  0\\
  0 & 0 & 1-\mathcal{V}^2 & 0\\
  2\mathcal{V}^4 &  0 &  0 & 1+\mathcal{V}^2
\end{bmatrix}.
\end{equation}

This state is of the same form as $\rho^{(0)}_{OBB}$, but accompanied by a prefactor of 1/2 success probability for successful fusion and a relabelling of $\mathcal{V}\mapsto\mathcal{V}^2$. This means we can describe the degradation of postselected $\rho^{(1)}_{OBB}$ post-fusion states through using lower visibility photons in the HBSG circuit.

For RS (two Schmidt) photons, the output state from a single HBSG of the type 4P8M is

\begin{equation}
\rho^{(0)}_{RS}=\frac{1}{4} \begin{bmatrix}
 1+\mathcal{V} & 0 &  0 &  \mathcal{V}^2+Y\\
  0 & 1-\mathcal{V} &  \mathcal{V}^2-Y\ &  0\\
  0 & \mathcal{V}^2-Y & 1-\mathcal{V} & 0\\
  \mathcal{V}^2+Y &  0 &  0 & 1+\mathcal{V}
\end{bmatrix}.
\end{equation}

We again assume each of the input photons is identical with pairwise visibility $\mathcal{V}$. The internal RS photon structure is $\varrho={p_0}\ket{k_0}\bra{k_0}+(1-{p_0})\ket{k_1}\bra{k_1}$. We have let $Y=\Tr(\varrho^4)={p_0}^4+(1-{p_0})^4$. Performing Type-II fusion on two copies of $\rho^{(0)}_{RS}$ preserves terms of $\rho^{(0)}_{RS}\otimes{\rho^{(0)}_{RS}}$ as

\begin{equation}
\begin{gathered}
\frac{1}{16}((\alpha\ket{00}\bra{00}+\alpha\ket{11}\bra{11}+\beta\ket{01}\bra{01}+\beta\ket{10}\bra{10}+\gamma\ket{00}\bra{11}+\gamma\ket{11}\bra{00}+\delta\ket{01}\bra{10}+\delta\ket{10}\bra{01})\\\otimes\\{(\alpha\ket{00}\bra{00}+\alpha\ket{11}\bra{11}+\beta\ket{01}\bra{01}+\beta\ket{10}\bra{10}+\gamma\ket{00}\bra{11}+\gamma\ket{11}\bra{00}+\delta\ket{01}\bra{10}+\delta\ket{10}\bra{01}))}.
\end{gathered}
\end{equation}

Here we simplify coefficients as $\alpha=1+\mathcal{V}$, $\beta=1-\mathcal{V}$, $\gamma=\mathcal{V}^2+Y$ and $\delta=\mathcal{V}^2-Y$. Expanding and grouping terms leads to

\begin{equation}
\begin{gathered}
\frac{1}{16}((\alpha^2+\beta^2)(\ket{00}\bra{00}+\ket{11}\bra{11})+2\alpha\beta(\ket{01}\bra{01}+\ket{10}\bra{10})\\+(\gamma^2+\delta^2)(\ket{00}\bra{11}+\ket{11}\bra{00})+2\delta\gamma(\ket{01}\bra{10}+\ket{10}\bra{01})).
\end{gathered}
\end{equation}

Redefining each coefficient in terms of $\mathcal{V}$ and $Y$ results in

\begin{equation}
\rho_{RS}^{(1)}=\frac{1}{4}\begin{bmatrix}
 1+\mathcal{V}^2 & 0 &  0 &  \mathcal{V}^4+Y^2\\
  0 & 1-\mathcal{V}^2 &  \mathcal{V}^4-Y^2 &  0\\
  0 & \mathcal{V}^4-Y^2 & 1-\mathcal{V}^2 & 0\\
  \mathcal{V}^4+Y^2 &  0 &  0 & 1+\mathcal{V}^2
\end{bmatrix}.
\end{equation}

This state is of the same form as $\rho^{(0)}_{RS}$ with $1/2$ successful fusion probability and a relabelling of $\mathcal{V}\mapsto\mathcal{V}^2$ and $Y\mapsto{Y^2}$. Remapping the initial RS photons used for the 4P8M HBSG with parameters $\mathcal{V}^{\prime}, {Y}^{\prime}$ requires

\begin{equation}
    \mathcal{V}^{\prime}={q_0}^2+(1-{q_0})^2\equiv{\mathcal{V}^2}=({p_0}^2+(1-{p_0})^2)^2,
\end{equation}
\begin{equation}
    Y^{\prime}={q_0}^4+(1-{q_0})^4\equiv{Y^2}=({p_0}^4+(1-{p_0})^4)^2.
\end{equation}

The only physical values which satisfy these mappings simultaneously are $p_0=1,q_0=1$ and $p_0=0,q_0=0$ which are the noiseless cases. This leads to the expectation of not being able to perfectly reconstruct $\rho^{(1)}_{RS}$ from a HBSG with partially distinguishable RS (two Schmidt) input photons as discussed in Sec. \ref{bosonicleakagecont} and modelled in Fig. \ref{fig:fusionresults}b, when performing postselection at the post-HBSG stage.

\section{Non-computational leakage in Schur states}\label{leakage-app}
\begin{figure}[!htbp]
	\centering
	\includegraphics[width=0.85\textwidth]{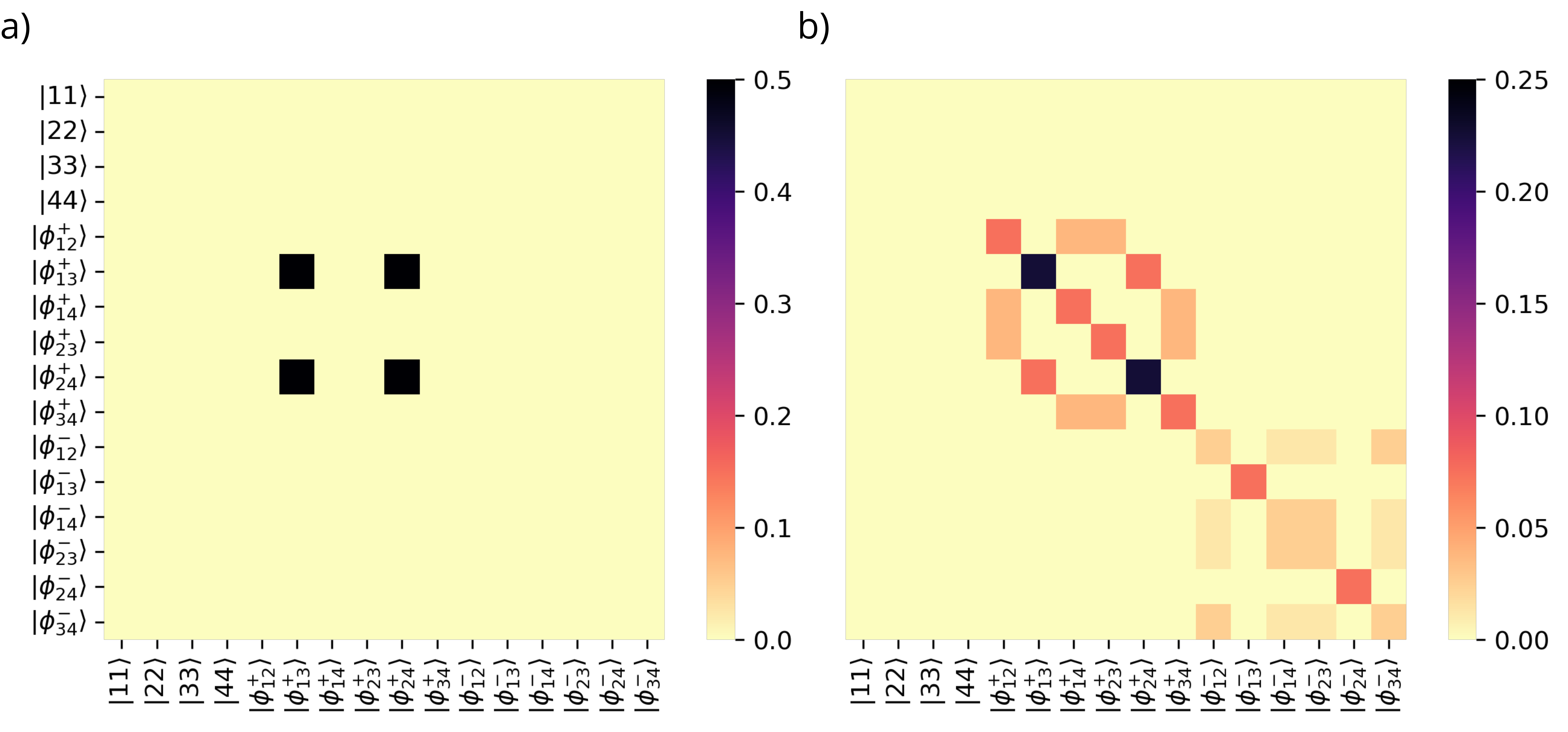}
	\caption{\label{fig:firstqrsmatrix}Reduced logical $\tilde{\rho}$ Schur states for the 4P8M HBSG (RS two Schmidt) with $\mathcal{V}=1,0.5$ in \textbf{a,b)} respectively. In the presence of partial distinguishability, leakage into the anti-symmetric two-photon subspace (lower right $6\times6$ sector) is seen. This subspace consists of both computational and non-computational terms with non-zero occupation for non-ideal photons.}
\end{figure}

To construct states $\tilde{\rho}$ attributed to Bell states, we require the Schur-Weyl transform on two ququarts (two dual-rail encoded qubits). The first quantized basis states $\{\ket{i,j},\ket{\phi^{\pm}_{i,j}}\}$ form the two-photon system space with a symmetric $10\times10$ sector (upper left) and $6\times6$ anti-symmetric sector (lower right). Here the ket labels $i,j$ indicate the spatial mode occupied by each photon, with computational and non-computational terms distributed across both the symmetric and anti-symmetric subspaces. Logical state degradation when characterizing states in the Schur-Weyl basis can be attributed to non-zero occupation of Schur-Weyl basis states which don't comprise the $\ket{\Phi^+}$ state. For ideal HBSG operation $\ket{\Phi^+}=\frac{1}{\sqrt{2}}(\ket{\phi^{+}_{1,3}}+\ket{\phi^{+}_{2,4}})$ (see Fig. \ref{fig:firstqrsmatrix}a). When $\mathcal{V}<1$  (RS two Schmidt) we have non-zero occupation in the anti-symmetric sector, both in computational states such as $\{\ket{\phi^{-}_{1,3}}, \ket{\phi^{-}_{2,4}}\}$ and non-computational states where two photons occupy the same dual-rail qubit modes, $\{\ket{\phi^{-}_{1,2}}, \ket{\phi^{-}_{3,4}}\}$ (see Fig. \ref{fig:firstqrsmatrix}b). The latter terms contribute to the leakage of $\tilde{\rho}$ out of the computational subspace and aren't captured by standard postselection. Whilst we only show $\tilde{\rho}$ matrices for RS photons here for concision, similar leakage mechanisms are exhibited for $\mathcal{V}<1$ OBB photons.

\end{document}